\documentclass[12pt]{article}

\usepackage{hyperref}
\hypersetup{
    colorlinks=true,
    linkcolor=blue,
    citecolor=blue,
    filecolor=magenta,
    urlcolor=green,
}
\usepackage{xcolor}
\usepackage{natbib}
\usepackage{graphicx}
\usepackage{multirow}
\usepackage{enumitem}
\usepackage{amsmath,amsthm,amssymb}
\usepackage{authblk}
\usepackage{makecell}
\usepackage{subcaption}
\usepackage{float}
\usepackage{placeins}
\usepackage{xparse}
\usepackage{etoolbox}
\usepackage[linesnumbered,ruled,vlined]{algorithm2e}

\def\real{\mathbb{R}}
\def\f{Fr\'echet }

\DeclareMathOperator*{\argmin}{arg\,min}

\newcommand{\Eo}{E_\oplus}
\def\T{{ \mathrm{\scriptscriptstyle T} }}

\newcommand{\figalt}[1]{\par\smallskip\noindent Alt text: #1\par}

\def\spacingset#1{\renewcommand{\baselinestretch}{#1}\small\normalsize}
\newcommand{\single}{\spacingset{1}}
\newcommand{\double}{\spacingset{1.9}}

\newcommand{\figuresize}[1]{\def\gddfigurewidth{#1\textwidth}}
\NewDocumentCommand{\figurebox}{m m m O{}}{\centering\includegraphics[width=\gddfigurewidth]{#4}}
\newsavebox{\gddtablebox}
\newcommand{\tbl}[2]{%
    \centering\caption{#1}%
    \sbox{\gddtablebox}{#2}%
    \ifdim\wd\gddtablebox>\linewidth
        \resizebox{\linewidth}{!}{\usebox{\gddtablebox}}%
    \else
        \usebox{\gddtablebox}%
    \fi}
\newenvironment{keywords}{\par\noindent\textit{Keywords:} }{\par}
\AtBeginEnvironment{figure}{\single\centering}
\AtBeginEnvironment{table}{\single\small}
\AtBeginEnvironment{algorithm}{\single}

\newtheorem{proposition}{Proposition}
\newtheorem{theorem}{Theorem}
\newtheorem{corollary}{Corollary}
\newtheorem{assumption}{Assumption}
\newtheorem{definition}{Definition}
\newtheorem{remark}{Remark}
\newtheorem{example}{Example}

\begin{document}

\single
\title{\bfseries Geodesic Difference-in-Differences}
\author[1,*]{Yidong Zhou}
\author[2,*]{Daisuke Kurisu}
\author[3]{Taisuke Otsu}
\author[1]{Hans-Georg M\"uller}
\affil[1]{Department of Statistics, University of California, Davis,\\
One Shields Avenue, Davis, California 95616, U.S.A.}
\affil[2]{Faculty of Economics, The University of Tokyo,\\
7-3-1, Hongo, Bunkyo-ku, Tokyo 113-0033, Japan\\}
\affil[3]{Department of Economics, London School of Economics,\\
Houghton Street, London, WC2A 2AE, U.K.\\}
\maketitle
\renewcommand*{\thefootnote}{\fnsymbol{footnote}}
\footnotetext[1]{The first two authors contributed equally to this work.}

\begin{abstract}
Difference-in-differences (DID) is a widely used quasi-experimental design for causal inference, traditionally applied to scalar or Euclidean outcomes, while extensions to outcomes residing in non-Euclidean spaces remain limited. Existing methods for such outcomes have primarily focused on univariate distributions, leveraging linear operations in the space of quantile functions, but these approaches cannot be directly extended to outcomes in general metric spaces. In this paper, we propose geodesic DID, a novel DID framework for outcomes in uniquely geodesic metric spaces that admit a geodesic transport structure, including distributions, networks, and manifold-valued data. To address the absence of algebraic operations in these spaces, we use geodesics as proxies for differences and introduce the geodesic average treatment effect on the treated (ATT) as the causal estimand. We establish the identification of the geodesic ATT and derive the convergence rate of its sample versions, employing tools from metric geometry and empirical process theory. This framework is further extended to the case of staggered DID settings, allowing for multiple time periods and varying treatment timings. To illustrate the practical utility of geodesic DID, we analyze health impacts of the Soviet Union's collapse using age-at-death distributions and assess effects of U.S. electricity market liberalization on electricity generation compositions.
\end{abstract}


\vfill
\newpage
\double

\section{Introduction}
Difference-in-differences (DID) is widely used to study the causal effect of an intervention, such as a policy change, a new social or healthcare programme, or an unexpected event, by exploiting observational panel data. Under the assumption that average outcomes of control and treated groups in the absence of an intervention obey parallel trends over time, DID identifies the average treatment effect on the treated (ATT) by comparing average outcomes of those groups for pre- and post-intervention periods. Estimation and inference for DID models are typically conducted via two-way fixed effects regression. DID is applied in various fields and application areas, including economics \citep{angr:09}, epidemiology and public health \citep{wing:18}, management science \citep{anto:10}, marketing \citep{vari:16}, and political science \citep{keel:15}.

Since its beginnings \citep[e.g.,][]{card:94}, the methodology of DID has been extended in various directions. For example, \citet{athe:06} proposed the changes-in-changes approach to allow heterogeneous treatment effects and infer the counterfactual distribution for the treated group; see \citet{toro:24} for an extension to multivariate outcomes and \citet{biew:22} for an alternative approach using distribution regression. Other developments in the DID framework include  integration with instrumental variables \citep{ye:23} and adaptations to factorial designs \citep{xu:24}. The fundamental parallel trends assumption, which underpins the validity of DID estimates, has also been extensively examined \citep{sofe:16, RoSa:23}. For the case of staggered treatment adoption, \citet{CaSa21} provided key insights into identification and estimation, with additional contributions from \citet{ding:19} and \citet{ben:22}, and a comprehensive overview of the topic by \citet{roth:23}.

In modern data science, one increasingly encounters complex-structured data, such as networks \citep{mull:22:11}, distributions \citep{pete:22}, compositional data \citep{scea:11} or trees \citep{bill:01}. Such data are inherently non-Euclidean and can be viewed as elements of a metric space.  Addressing causal inference for metric space-valued data is a very recent development and includes doubly robust estimators for distributional outcomes \citep{lin:23}, for outcomes residing in general geodesic metric spaces \citep{kuri:24} and also the synthetic control method for distributional outcomes \citep{guns:23}. Almost all existing DID methodologies focus on scalar or Euclidean outcomes, with the exception of \citet{opok:23}, where causal effects on distributional outcomes are studied within the DID framework. This approach leverages restricted linear operations in the space of quantile functions and is limited to the special case of univariate distributions. Since it is based on representing distributions by quantile functions, it cannot be extended to more general data objects.

We introduce here geodesic DID, a novel framework for DID analysis of outcomes residing in a broad class of uniquely geodesic metric spaces that admit a geodesic transport structure. Such spaces include distributions, networks, compositional data, and manifold-valued objects that arise frequently in modern statistical applications. To address the absence of algebraic operations such as addition and scalar multiplication in these settings, we use geodesics as geometric proxies for differences and define the geodesic ATT as the causal estimand. We establish identification of the geodesic ATT and derive convergence rates for its empirical estimators using tools from metric geometry and empirical process theory. An extension to staggered DID settings, allowing for multiple time periods and variation in treatment timing, is provided in the Supplementary Material. An R implementation of geodesic DID, together with reproducibility materials for the simulation studies and data analyses, is available at \texttt{\url{https://github.com/yidongzhou/geodesic-did}}.

\section{Preliminaries on geodesic calculus}\label{sec:pre}
Let $(\mathcal{M},d)$ be a bounded and separable metric space. A \textit{curve} in $\mathcal{M}$ is a continuous map $\gamma:[a,b]\to\mathcal{M}$ with length
\[L(\gamma) = \sup\sum_{k=0}^{K-1} d\{\gamma(t_{k}),\gamma(t_{k+1})\},\]
where the supremum is taken over all $K\geq1$ and partitions $a = t_0 \leq t_1 \leq \cdots \leq t_K = b$ and the curve is rectifiable if $L(\gamma) < \infty$. Two curves $\gamma_1$ and $\gamma_2$ are said to be equivalent if $\gamma_1\circ \phi_1=\gamma_2\circ\phi_2$ for non-decreasing and continuous functions $\phi_1$ and $\phi_2$. In this case, $\gamma_1$ is said to be a reparametrization of $\gamma_2$ with $L(\gamma_1)=L(\gamma_2)$. A curve $\gamma : [a,b] \to \mathcal{M}$ is said to have constant speed if for all $a \leq s \leq t \leq b$,
\[L(\gamma_{[s,t]}) = \frac{t-s}{b-a} L(\gamma),\]
where $\gamma_{[s,t]}$ denotes the restriction of $\gamma$ to $[s,t]$.

Given $\alpha,\beta \in \mathcal{M}$, a curve $\gamma:[a,b]\to\mathcal{M}$ is said to connect $\alpha$ to $\beta$ if $\gamma(a) = \alpha$ and $\gamma(b) = \beta$. By construction of the length function $L$, $d(\alpha,\beta) \leq L(\gamma)$ for any curve $\gamma$ connecting $\alpha$ to $\beta$. The space $\mathcal{M}$ is called a \textit{length space} if, for all $\alpha,\beta \in \mathcal{M}$,
\begin{equation}
\label{eq:dinf}
d(\alpha, \beta) = \inf_{\gamma} L(\gamma),
\end{equation}
where the infimum is taken over all curves $\gamma$ connecting $\alpha$ to $\beta$. A length space is said to be a \textit{geodesic space} if, for all $\alpha,\beta \in \mathcal{M}$, the infimum on the right-hand side of \eqref{eq:dinf} is attained.

In a geodesic space, a \textit{geodesic} between $\alpha$ and $\beta$ is any constant speed reparametrization $\gamma:[0,1]\to\mathcal{M}$ of a curve attaining the infimum in \eqref{eq:dinf}. The geodesic connecting $\alpha$ and $\beta$ is denoted by $\gamma_{\alpha, \beta}$. If there exists only one such geodesic for all $\alpha, \beta \in \mathcal{M}$, the space $\mathcal{M}$ is referred to as a \textit{uniquely geodesic space} \citep{brids:99}.

For points $\alpha, \beta, \zeta \in \mathcal{M}$, we define the following operations on geodesics. The concatenation of two geodesics $\gamma_{\alpha, \zeta}$ and $\gamma_{\zeta, \beta}$ is given by $\gamma_{\alpha, \zeta} \oplus \gamma_{\zeta, \beta} = \gamma_{\alpha, \beta}$, while the reversal of a geodesic $\gamma_{\alpha, \beta}$ is defined as $\ominus\gamma_{\alpha, \beta} = \gamma_{\beta, \alpha}$. The difference between two geodesics with a common start point is defined as $\ominus\gamma_{\zeta, \alpha} \oplus \gamma_{\zeta, \beta}= \gamma_{\alpha, \beta}$. These operations provide geometric analogues of addition and subtraction in Euclidean space, without imposing a linear structure on $\mathcal{M}$. Such constructions also underpin previous work
on doubly robust estimation of average treatment effects in geodesic spaces \citep{kuri:24}.

Next, we present examples of uniquely geodesic spaces encountered in real-world applications. We use these example spaces to demonstrate the practical performance of the proposed approach in simulations and data analyses. Section~S.4 of the Supplementary Material provides conditions under which the assumptions and convergence rates in Section~\ref{sec:the} hold for these example spaces.

\begin{example}[Functional data]
\label{exm:fun}
Functional data consist of observations that are naturally viewed as real-valued functions defined over a domain $\mathcal{T}$, such as time or spatial location. Let $\mathcal{T}$ be a compact interval and consider the Hilbert space $L^2(\mathcal{T})$ of square-integrable functions equipped with the inner product $\langle f,g\rangle = \int_{\mathcal{T}} f(t)g(t)\,\mathrm{d}t$ and the induced metric
\[d_{L^2}(f,g)=\left[\int_{\mathcal{T}} \{f(t)-g(t)\}^2\,\mathrm{d}t\right]^{1/2}.\]
Under this metric, $L^2(\mathcal{T})$ is a uniquely geodesic space. The geodesic between two functions $f$ and $g$ is given by the linear interpolation $\gamma_{f,g}(t)=(1-t)f + t g$, $t \in [0,1]$.
\end{example}
\begin{example}[Univariate probability distributions]
\label{exm:upd}
Let $\mathcal{P}_2(\mathbb{R}^d)$ denote the space of Borel probability measures on $\mathbb{R}^d$ with finite second moments. Equipped with the $2$-Wasserstein metric,
\[d_{\mathcal{W}}^2(\mu_1, \mu_2)=\inf_{\pi \in \Pi(\mu_1,\mu_2)}\int_{\mathbb{R}^d\times\mathbb{R}^d}\|x-y\|^2\,\mathrm{d}\pi(x,y),\]
where $\Pi(\mu_1,\mu_2)$ denotes the set of couplings of $\mu_1$ and $\mu_2$, the space $(\mathcal{P}_2(\mathbb{R}^d),d_{\mathcal{W}})$ forms a geodesic metric space \citep{vill:09, pana:20}. In one dimension, this distance admits the explicit representation
\[d_{\mathcal{W}}^2(\mu_1, \mu_2)=\int_0^1\{ F_{\mu_1}^{-1}(p) - F_{\mu_2}^{-1}(p) \}^2\,\mathrm{d}p,\]
where $F_{\mu_1}^{-1}$ and $F_{\mu_2}^{-1}$ denote the quantile functions of $\mu_1$ and $\mu_2$, respectively. The geodesic between $\mu_1$ and $\mu_2$ under the $2$-Wasserstein metric is given by $\gamma_{\mu_1, \mu_2}(t) = (\mathrm{id} + t(F_{\mu_2}^{-1} \circ F_{\mu_1} - \mathrm{id}))_\#\mu_1$, where $\mathrm{id}$ denotes the identity map and $T_\#\mu_1$ denotes the pushforward of $\mu_1$ by $T$ \citep{mcca:97}. For the theoretical developments in Section~\ref{sec:the}, we focus on absolutely continuous probability measures supported on a closed interval on $\mathbb{R}$ with bounded and non-vanishing densities.
\end{example}
\begin{example}[Multivariate probability distributions]
\label{exm:mpd}
Computation of $d_{\mathcal{W}}$ in high dimensions can be demanding. Entropic regularization (Sinkhorn distances) \citep{cutu:13} and sliced Wasserstein distances \citep{bonn:15} are commonly used approximations to reduce computational cost. An alternative geometric structure is provided by the Fisher--Rao metric on the space of absolutely continuous distributions. Let $\mathcal{H}$ denote the separable Hilbert space of square-integrable functions with inner product $\langle g_1,g_2\rangle=\int g_1(x)g_2(x)\,\mathrm{d}x$. Under the square-root representation $f \mapsto \sqrt{f}$, densities are mapped to the positive orthant of the Hilbert sphere $\mathcal{S}_+=\{ g \in \mathcal{H} : \|g\|_{L^2}=1, g \ge 0 \}$. The Fisher--Rao distance between densities $f_1$ and $f_2$ is
\[d_{FR}(f_1,f_2)=\arccos\left(\langle \sqrt{f_1}, \sqrt{f_2} \rangle\right),\]
which coincides with the spherical geodesic distance between their square roots. Geodesics under this metric are great-circle arcs,
\[\gamma_{f_1, f_2}(t)=\frac{\sin((1-t)\theta)}{\sin\theta}\,\sqrt{f_1}+\frac{\sin(t\theta)}{\sin\theta}\,\sqrt{f_2},\quad t \in [0,1],\]
where $\theta=\arccos(\langle \sqrt{f_1}, \sqrt{f_2}\rangle)$, and the corresponding density path is obtained by squaring $\gamma(t)$. This representation yields an explicit and computationally convenient geodesic structure suitable for high-dimensional distributional data. The choice between these geometries is substantive and should reflect the scientific meaning of distributional change. The Fisher--Rao metric used in the multivariate simulation provides one computationally convenient illustration and is not intended as a general recommendation over the $2$-Wasserstein metric. For the theoretical developments in Section~\ref{sec:the} under the $2$-Wasserstein metric, we focus on multivariate Gaussian distributions with bounded mean vectors and covariance eigenvalues bounded above and away from zero.
\end{example}
\begin{example}[Networks]\label{exm:net}
Network data describe relationships among entities and are commonly represented by graph Laplacian matrices \citep{kola:14,mull:22:11,seve:22}. Consider the collection $\mathcal{L}$ of Laplacian matrices corresponding to undirected, weighted networks with $m$ nodes and bounded edge weights. Equipping $\mathcal{L}$ with the Frobenius metric $d_F(L_1,L_2)=\|L_1-L_2\|_F$,
induces a geodesic structure under which the geodesic between $L_1$ and $L_2$ is given by the straight-line interpolation $\gamma_{L_1,L_2}(t)=(1-t)L_1+tL_2$, $t\in[0,1]$.
\end{example}
\begin{example}[Symmetric positive-definite matrices]\label{exm:spd}
Symmetric positive-definite (SPD) matrices arise in applications such as covariance modelling and diffusion tensor imaging \citep{dryd:09,huan:17}. Let $\mathrm{Sym}_m^+$ denote the set of $m\times m$ SPD matrices. This set is an open subset of the space of symmetric matrices and can be equipped with several geodesic metrics. Under the Frobenius metric, the geometry is flat and the geodesic between $A$ and $B$ is the linear path $\gamma_{A,B}(t)=(1-t)A+tB$, $t\in[0,1]$. Under the Log-Euclidean metric \citep{arsi:07}, $d_{LE}(A,B)=\|\log A-\log B\|_F$, geodesics are obtained by linear interpolation in the matrix logarithm domain, $\gamma_{A,B}(t)=\exp\left((1-t)\log A+t\log B\right)$. Equipping $\mathrm{Sym}_m^+$ with the affine-invariant Riemannian metric \citep{penn:06}, $d_{AI}(A, B)=\|\log(A^{-1/2}BA^{-1/2})\|_F$, yields a curved Riemannian geometry, with geodesics
\[\gamma_{A,B}(t)=A^{1/2}\exp\left(t\,\log(A^{-1/2}BA^{-1/2})\right)A^{1/2},\quad t\in[0,1].\]
Other metrics such as the power metric \citep{dryd:09} and Log-Cholesky metric \citep{lin:19:1} also endow $\mathrm{Sym}_m^+$ with a uniquely geodesic structure.
\end{example}
\begin{example}[Compositional data]
\label{exm:cd}
Compositional data, which are represented as proportions summing to 1, reside in the simplex
\[\Delta^{d-1}=\left\{y\in\mathbb{R}^d:y_j\geq0,\ j=1,\ldots,d,\ \text{and }\sum_{j=1}^d y_j=1\right\}.\]
Through the component-wise square-root transformation $\sqrt{y}=(\sqrt{y_1},\ldots,\sqrt{y_d})^\T$, the simplex is mapped to the first orthant $\mathcal{S}_+^{d-1}=\{z\in\mathcal{S}^{d-1}:z_j\geq0,\ j=1,\ldots,d\}$ of the unit sphere \citep{scea:11,scea:14}. Equipping $\mathcal{S}_+^{d-1}$ with the Riemannian metric $d_S(z_1,z_2)=\arccos(z_1^\T z_2)$ induces a uniquely geodesic structure. For $\theta=\arccos(z_1^\T z_2)$, the geodesic from $z_1$ to $z_2$ is
\[\gamma_{z_1,z_2}(t)=\frac{\sin((1-t)\theta)}{\sin\theta}\,z_1+\frac{\sin(t\theta)}{\sin\theta}\,z_2,
\quad t\in[0,1].\]
\end{example}

\section{Geodesic difference-in-differences}\label{sec:met}
Consider the canonical DID model with Euclidean outcomes, two periods $t\in\{0,1\}$, and units $i=1,\ldots,n$. Let $D_i$ be the group indicator, where $D_i=1$ if unit $i$ belongs to the treated group and $D_i=0$ otherwise. For unit $i$ at time $t$, let $D_{i,t}$ be the treatment status indicator and let $Y_{i,t}(0)$ and $Y_{i,t}(1)$ denote the potential outcomes without and with treatment, respectively. The observed outcome is $Y_{i,t}=D_{i,t}Y_{i,t}(1)+(1-D_{i,t})Y_{i,t}(0)$. This potential-outcomes formulation encodes the stable unit treatment value assumption: unit $i$'s outcome does not depend on the treatment status of any other unit $j\ne i$, ruling out spillovers and general equilibrium effects \citep{roth:23}. We assume that treated units receive treatment between the two periods, whereas control units remain untreated in both periods. Thus, $(D_{i,0},D_{i,1})=(0,0)$ if $D_i=0$, and $(D_{i,0},D_{i,1})=(0,1)$ if $D_i=1$.

The causal estimand of interest is the \textit{average treatment effect on the treated} (ATT), $\tau=E\{Y_{i, 1}(1)-Y_{i, 1}(0)\mid D_i=1\}$. Then under the parallel trends assumption that
\[E\{Y_{i, 1}(0)-Y_{i, 0}(0)\mid D_i=1\}=E\{Y_{i, 1}(0)-Y_{i, 0}(0)\mid D_i=0\},\]
the ATT is identified by the canonical DID estimand
\begin{equation}\label{eq:cdid}
    \tau=E\{Y_{i, 1}-Y_{i, 0}\mid D_i=1\}-E\{Y_{i, 1}-Y_{i, 0}\mid D_i=0\}.
\end{equation}

We now allow $Y_{i,t}$ to be a random object in a uniquely geodesic space $(\mathcal{M},d)$, retaining the same treatment timing and SUTVA. Thus, $Y_{i,t}=Y_{i,t}(0)$ if $D_{i,t}=0$ and $Y_{i,t}=Y_{i,t}(1)$ if $D_{i,t}=1$. We assume that $\{(Y_{i,0},Y_{i,1},D_i)\}_{i=1}^n$ is i.i.d. from a joint distribution $P$, and write $(Y_0,Y_1,D)\sim P$ for a generic observation in $\mathcal{M}\times\mathcal{M}\times\{0,1\}$.

For a random object $Y\in\mathcal{M}$, the \f mean of $Y$, extending the usual notion of mean, is
\[\Eo\{Y\}=\argmin_{\omega\in\mathcal{M}}E\{d^2(Y, \omega)\},\]
where the existence and uniqueness of the minimizer are guaranteed for Hadamard spaces \citep{stur:03} and the example spaces described in Examples~\ref{exm:fun}--\ref{exm:cd}. The conditional \f mean of $Y$ given $X\in\mathbb{R}^p$ \citep{mull:19:6} is analogously defined as
\[\Eo\{Y\mid X\}=\argmin_{\omega\in\mathcal{M}}E\{d^2(Y, \omega)\mid X\}.\]

To obtain the DID estimand for outcomes in $\mathcal{M}$, it is necessary to define a notion of subtraction in the uniquely geodesic space $\mathcal{M}$. Following \citet{kuri:24}, we represent the difference between two points $\alpha,\beta\in\mathcal{M}$ by the geodesic $\gamma_{\alpha,\beta}$. In the canonical DID estimand \eqref{eq:cdid}, a second difference is computed between two differences for the treated and control groups. Extending this concept to $\mathcal{M}$ requires quantifying differences between geodesics, which can be achieved with geodesic transport maps under the following assumption \citep{zhu:23}, which has been referred to as ubiquity of geodesics.
\begin{assumption}\label{asp:ug}
Let $(\mathcal{M}, d)$ be a uniquely geodesic space. For any two points $\alpha, \beta \in \mathcal{M}$, there exists a \textit{geodesic transport map} $\Gamma_{\alpha,\beta}:\mathcal{M}\to\mathcal{M}$ with the following property: $\Gamma_{\alpha,\beta}(\alpha)=\beta$ and for any $\omega \in \mathcal{M}$, there exists a unique point $\zeta \in \mathcal{M}$ such that $\Gamma_{\alpha, \beta}(\omega) = \zeta$.\end{assumption}

Assumption~\ref{asp:ug} formalizes the idea that the displacement along the geodesic from $\alpha$ to $\beta$ can be replicated at any other point in the space. Specifically, $\Gamma_{\alpha,\beta}$ moves an arbitrary base point $\omega$ in a manner that mirrors the movement from $\alpha$ to $\beta$, producing the endpoint $\zeta=\Gamma_{\alpha,\beta}(\omega)$. Thus, the geodesic from $\alpha$ to $\beta$ induces a transport operation that can be applied at any base point, analogous to translating a displacement vector in Euclidean geometry. In $\mathbb R^p$, $\Gamma_{\alpha,\beta}(\omega)=\omega+\beta-\alpha$. The same translation structure applies to arbitrary vector spaces, including the space $L^2(\mathcal T)$ of functional data in Example~\ref{exm:fun}.

The assumption is also satisfied in several nonlinear metric spaces commonly encountered in applications. In the univariate $2$-Wasserstein space of Example~\ref{exm:upd}, geodesics are given by displacement interpolation, and the transport construction is induced by the optimal transport map. For multivariate Gaussian distributions under the $2$-Wasserstein metric (Example~\ref{exm:mpd}), the transport construction is induced by the affine optimal transport map between Gaussian measures. For SPD matrices equipped with Riemannian metrics such as the affine-invariant metric (Example~\ref{exm:spd}) and for compositional data endowed with the spherical metric (Example~\ref{exm:cd}), the spaces admit a Riemannian manifold structure. In these settings, the transport map is constructed by transporting tangent vectors via parallel transport \citep{lee:18} and mapping them back to the manifold using the exponential map. A further example is provided by the Billera--Holmes--Vogtmann tree space of phylogenetic trees \citep{bill:01}, described in Section~S.2 of the Supplementary Material. This space is globally non-positively curved and uniquely geodesic, and Assumption~\ref{asp:ug} holds in particular for trees with four leaves.

In general metric spaces, it is conceivable that different constructions satisfying Assumption~\ref{asp:ug} could be adopted. In such cases, the proposed framework is understood relative to the chosen transport structure. In the spaces considered in this paper, the underlying geometry naturally determines a standard and widely accepted transport construction, as described for the examples in Section~\ref{sec:pre}, with detailed constructions provided in Section~S.3 of the Supplementary Material.

\begin{remark}[Smooth Riemannian manifolds]
On a smooth Riemannian manifold \citep{lee:18}, geodesic transport can be constructed using logarithmic and exponential maps together with parallel transport \citep{yuan:12,chen:23:1}. Let $T_\omega\mathcal M$ be the tangent space at $\omega$. For $\alpha$ and $\beta$ in a normal neighbourhood, $\mathrm{Log}_\alpha(\beta)\in T_\alpha\mathcal M$ is the initial velocity of the unique geodesic from $\alpha$ to $\beta$. Given another base point $\omega$, let $\mathcal P_{\alpha\to\omega}$ transport this tangent vector from $T_\alpha\mathcal M$ to $T_\omega\mathcal M$ along the minimizing geodesic from $\alpha$ to $\omega$, and define
$\Gamma_{\alpha,\beta}(\omega)=\mathrm{Exp}_\omega\{\mathcal P_{\alpha\to\omega}(\mathrm{Log}_\alpha(\beta))\}$. This construction satisfies Assumption~\ref{asp:ug} whenever the exponential map is defined on the transported vector and the relevant minimizing geodesics are unique.
\end{remark}

We now extend the notion of subtraction to geodesics with different starting points by aligning the subtrahend with the minuend using the geodesic transport map.
\begin{definition}\label{def:geo}
    For any points $\alpha, \beta, \omega, \zeta\in\mathcal{M}$ with $\alpha\neq\omega$, the difference between two geodesics is defined as
    \[\ominus\gamma_{\alpha, \beta}\oplus\gamma_{\omega, \zeta}=\ominus\gamma_{\omega, \zeta'}\oplus\gamma_{\omega, \zeta}=\gamma_{\zeta', \zeta},\]
    where $\zeta'=\Gamma_{\alpha, \beta}(\omega)$.
\end{definition}
The above operation is straightforward in the vector space $\mathbb{R}^p$, where $\zeta'=\omega+\beta-\alpha$ and we verify that $\zeta-\zeta'=-(\beta-\alpha)+(\zeta-\omega)$.

Let $\nu_{d, t}(0)=\Eo\{Y_{i, t}(0)\mid D_i=d\}$ and $\nu_{d, t}(1)=\Eo\{Y_{i, t}(1)\mid D_i=d\}$ for $d, t\in \{0, 1\}$. We introduce the following notions of causal effects in uniquely geodesic spaces.

\begin{definition}[Geodesic ATT]
    The geodesic average treatment effect on the treated (GATT) $\tau$ is defined as the geodesic connecting $\nu_{1,1}(0)$ and $\nu_{1,1}(1)$, i.e. $\tau=\gamma_{\nu_{1,1}(0), \nu_{1,1}(1)}$.
\end{definition}

The GATT is a geodesic, with its direction and length capturing the direction and magnitude of the causal effect when outcomes are random objects in uniquely geodesic spaces.

\begin{assumption}\label{asp:pt}
$\Gamma_{\nu_{0, 0}(0), \nu_{0, 1}(0)}(\nu_{1, 0}(0))=\nu_{1, 1}(0)$.
\end{assumption}

\begin{figure}
    \figuresize{0.6}
	\figurebox{20pc}{}{}[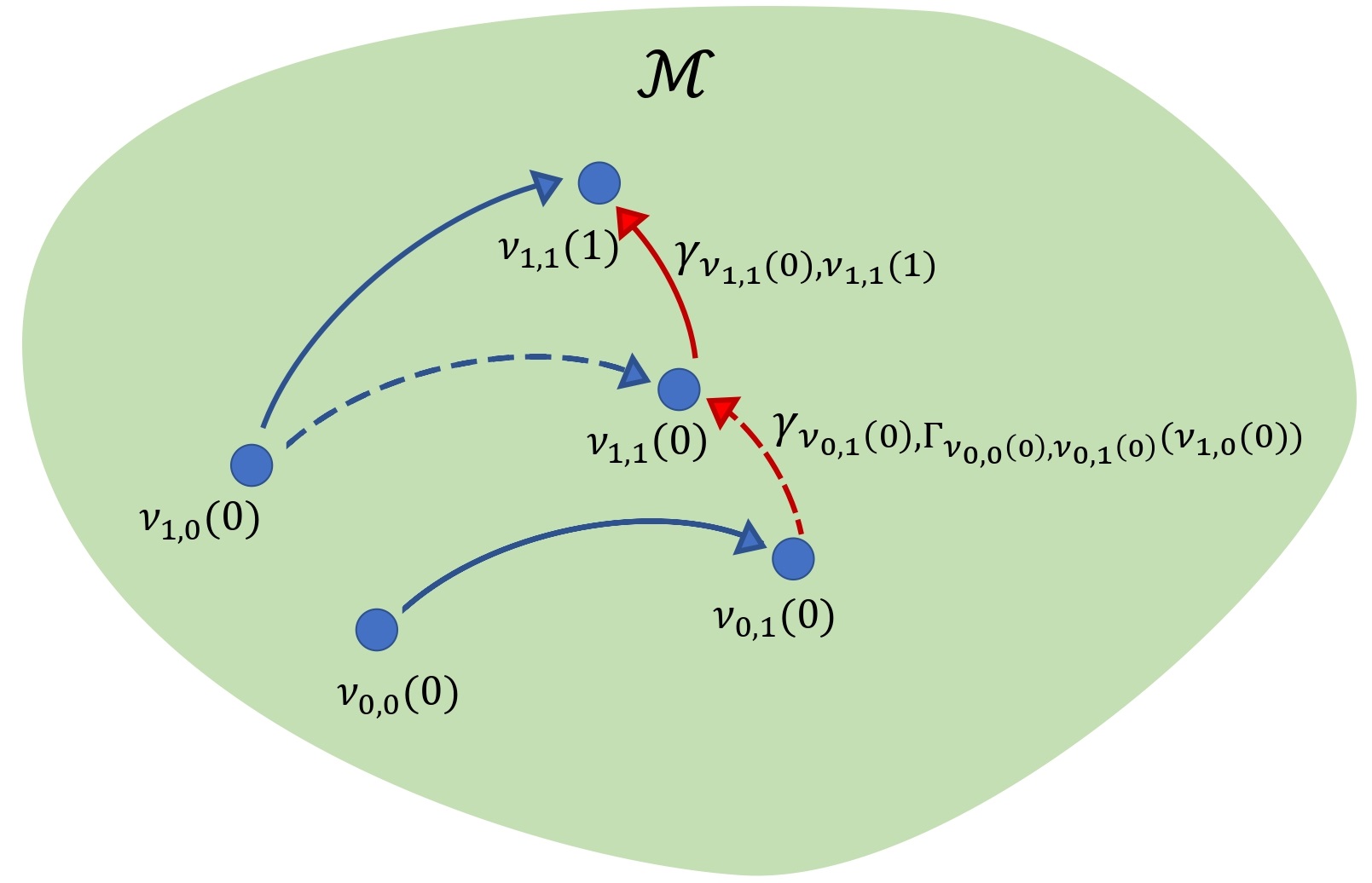]
    \caption{Illustration of geodesic DID. The blue solid circles symbolize objects in the uniquely geodesic space $\mathcal{M}$. The arrows emanating from the circles represent the geodesic paths connecting them. The red arrows represent the geodesic transport from $\nu_{0,1}(0)$ to $\nu_{1,1}(0)$ and the proposed GATT.}
    \figalt{Five points in $\mathcal M$ show the observed group-time means, the transported control-group change ending at the treated counterfactual mean $\nu_{1,1}(0)$, and the GATT from $\nu_{1,1}(0)$ to $\nu_{1,1}(1)$.}
    \label{fig:GDD-image}
\end{figure}

Assumption~\ref{asp:pt} is a direct generalization of the parallel trends assumption for the case where responses reside in a uniquely geodesic space. It asserts that in the absence of treatment, the geodesic connecting the control group's outcomes in the pre- and post-treatment periods, when extended from the treated group's pre-treatment outcome, should end at the treated group's post-treatment outcome. This ensures that, without treatment, the trajectory of outcomes for the treated group aligns with the geodesic trend of the control group; see Figure~\ref{fig:GDD-image} for an illustration. In the case of Euclidean outcomes, Assumption~\ref{asp:pt} reduces to $\nu_{1, 0}(0)+\{\nu_{0, 1}(0)-\nu_{0, 0}(0)\}=\nu_{1, 1}(0)$, which corresponds to the standard parallel trends assumption in the canonical DID framework.

Writing $\nu_{d, t}=\Eo\{Y_{i, t}\mid D_i=d\}$ for $d, t\in \{0, 1\}$, the GATT can be identified as follows.
\begin{theorem}\label{thm:GDD-id}
Suppose that Assumptions~\ref{asp:ug} and~\ref{asp:pt} hold. Then we have
\begin{equation}\label{eq:tau}
    \tau = \ominus \gamma_{\nu_{0,0},\nu_{0,1}} \oplus \gamma_{\nu_{1,0},\nu_{1,1}}=\gamma_{\nu_{1,1}', \nu_{1,1}},
\end{equation}
where $\ominus \gamma_{\nu_{0,0},\nu_{0,1}} \oplus \gamma_{\nu_{1,0},\nu_{1,1}}$ is the geodesic DID estimand and $\nu_{1, 1}'=\Gamma_{\nu_{0,0}, \nu_{0,1}}(\nu_{1,0})$.
\end{theorem}

In the Euclidean space $(\mathcal{M},d)=(\mathbb R,\|\cdot\|)$, $\gamma_{\alpha,\beta}(t)=\alpha+t(\beta-\alpha)$ and $\Gamma_{\alpha,\beta}(\omega)=\omega+\beta-\alpha$. Hence the geodesic DID estimand starts at $\nu_{1,0}+(\nu_{0,1}-\nu_{0,0})$ and ends at $\nu_{1,1}$. Under parallel trends, its signed displacement is
\begin{align*}
&\nu_{1,1} - \{\nu_{1,0}+(\nu_{0,1} - \nu_{0,0})\}
= (\nu_{1,1} - \nu_{1,0}) - (\nu_{0,1} - \nu_{0,0})\\
&= E\{Y_{i, 1}-Y_{i, 0}\mid D_i=1\}-E\{Y_{i, 1}-Y_{i, 0}\mid D_i=0\}
= E\{Y_{i, 1}(1)-Y_{i, 1}(0)\mid D_i=1\}.
\end{align*}
Therefore, Theorem~\ref{thm:GDD-id} extends the Euclidean ATT $\tau=E\{Y_{i,1}(1)-Y_{i,1}(0)\mid D_i=1\}$ to uniquely geodesic spaces.

\section{Estimation and theory}\label{sec:the}

Given a sample $\{(Y_{i, 0}, Y_{i, 1}, D_i)\}_{i=1}^n$, the geodesic DID estimand of the GATT, namely $\tau=\gamma_{\nu_{1,1}', \nu_{1,1}}$, is estimated by replacing the \f means with their sample counterparts \begin{equation}\label{eq:tauhat}
\hat{\tau}=\gamma_{\hat{\nu}_{1,1}', \hat{\nu}_{1,1}},\quad\hat{\nu}_{1,1}'=\Gamma_{\hat{\nu}_{0,0},\hat{\nu}_{0,1}}(\hat{\nu}_{1,0}),
\end{equation}
where for $t \in \{0,1\}$,
\begin{align*}
\hat{\nu}_{0,t} &= \argmin_{\omega \in \mathcal{M}}{\frac{1}{n_0}}\sum_{i=1}^n (1-D_i)d^2(Y_{i,t},\omega),\ n_0=\sum_{i=1}^n (1-D_i),\\
\hat{\nu}_{1,t} &= \argmin_{\omega \in \mathcal{M}}{\frac{1}{n_1}}\sum_{i=1}^n D_id^2(Y_{i,t},\omega),\  n_1=\sum_{i=1}^n D_i.
\end{align*}

Our goal is to establish the consistency of $\hat{\tau}$ in \eqref{eq:tauhat} and to derive its convergence rate. Since $\tau$ is a geodesic, a properly defined metric on the space of geodesics is required to evaluate the performance of $\hat{\tau}$. We define the space of geodesics as follows. 

\begin{definition}
    The space of geodesics in the uniquely geodesic space $(\mathcal{M}, d)$ is
    \[\mathcal{G}(\mathcal{M})=\{\gamma_{\alpha, \beta}:\alpha, \beta\in\mathcal{M}\}.\]
\end{definition}

To compare geodesics, we define a binary relation $\sim$ on $\mathcal{G}(\mathcal{M})$ through the geodesic transport map, where  $\gamma_{\alpha_1, \beta_1}\sim\gamma_{\alpha_2, \beta_2}$ if and only if $\Gamma_{\alpha_1, \beta_1}(\omega)=\Gamma_{\alpha_2, \beta_2}(\omega)$ for all $\omega\in\mathcal{M}$.

\begin{proposition}\label{prop:equi}
The relation $\sim$ is an equivalence relation on $\mathcal{G}(\mathcal{M})$.
\end{proposition}

This equivalence relation, drawing on the ubiquity Assumption~\ref{asp:ug}, partitions the space of geodesics $\mathcal{G}(\mathcal{M})$ into disjoint equivalence classes. The equivalence class of $\gamma_{\alpha, \beta}\in\mathcal{G}(\mathcal{M})$ is denoted by $[\gamma_{\alpha, \beta}]$. The set of all equivalence classes $\mathcal{G}(\mathcal{M})/\sim=\{[\gamma_{\alpha, \beta}]: \gamma_{\alpha, \beta}\in\mathcal{G}(\mathcal{M})\}$ is the quotient space of $\mathcal{G}(\mathcal{M})$, for which  we define a metric as follows.  For any $[\gamma_{\alpha_1, \beta_1}], [\gamma_{\alpha_2, \beta_2}]\in\mathcal{G}(\mathcal{M})/\sim$, let
\begin{equation}\label{eq:dg}
    d_{\mathcal{G}}([\gamma_{\alpha_1, \beta_1}], [\gamma_{\alpha_2, \beta_2}])=d_{\mathcal{G},\omega_\oplus}([\gamma_{\alpha_1, \beta_1}], [\gamma_{\alpha_2, \beta_2}])=d(\Gamma_{\alpha_1, \beta_1}(\omega_\oplus), \Gamma_{\alpha_2, \beta_2}(\omega_\oplus)),
\end{equation}
where $\omega_\oplus\in\mathcal{M}$ is an arbitrary reference point, chosen e.g., as
the \f mean of the outcomes.

\begin{proposition}\label{prop:geo}
    $d_{\mathcal{G}}$ is a metric on the quotient space $\mathcal{G}(\mathcal{M})/\sim$.
\end{proposition}

With a slight abuse of notation, we write $d_{\mathcal G}(\gamma_1,\gamma_2)$ for $d_{\mathcal G}([\gamma_1],[\gamma_2])$. $(\mathcal{G}(\mathcal{M})/\sim, d_{\mathcal{G}})$ is thus a metric space, and the estimation error of $\hat{\tau}$ can be quantified by $d_{\mathcal{G}}(\hat{\tau}, \tau)$, where the reference point is chosen as $\nu_{1, 1}'=\Gamma_{\nu_{0,0}, \nu_{0,1}}(\nu_{1,0})$. This metric compares two geodesic changes after aligning them at a common reference point. The chosen reference is the treated group's counterfactual post-treatment mean, so the distance directly measures discrepancy between the estimated and population causal changes at the relevant treated-group starting point. We require the following regularity assumptions.

\begin{assumption}\label{asp:iso}
    There exists a constant $C_1>0$ such that for any  $\alpha, \beta, \omega, \zeta \in\mathcal{M}$,
$d(\Gamma_{\alpha, \beta}(\omega), \Gamma_{\alpha, \beta}(\zeta))\leq C_1d(\omega,\zeta)$.
\end{assumption}
\begin{assumption}\label{asp:com}
There exists a constant $C_2>0$ such that for any  $\alpha_1,\alpha_2, \beta_1,\beta_2, \omega \in\mathcal{M}$,
$d(\Gamma_{\alpha_1, \beta_1}(\omega), \Gamma_{\alpha_2, \beta_2}(\omega))\leq C_2\{d(\alpha_1, \alpha_2)+d(\beta_1, \beta_2)\}$.
\end{assumption}
Assumption~\ref{asp:iso} requires the geodesic transport map to be Lipschitz continuous.  Assumption~\ref{asp:com} ensures that for any two geodesics, when moved to the same starting point, the distance between their new endpoints is bounded by the sum of the distances between their original starting points and endpoints. Together, these conditions propagate estimation errors in the group-time \f means through the transport operation: Assumption~\ref{asp:iso} controls changes in the base point, while Assumption~\ref{asp:com} controls changes in the transported geodesic. Both assumptions are trivially satisfied in Euclidean spaces with $C_1=C_2=1$.
\begin{assumption}\label{asp:fm}
For $d, t\in\{0, 1\}$, the following conditions hold.

(i) The population \f mean $\nu_{d,t}$ and its empirical counterpart $\hat{\nu}_{d,t}$ exist and are unique, the latter almost surely, and, for any $\varepsilon>0$, $\inf_{d(\nu_{d,t},\omega)>\varepsilon}E\{d^2(Y_t,\omega)\mid D=d\}>E\{d^2(Y_t,\nu_{d,t})\mid D=d\}$.

(ii) Let $B_\delta(\nu_{d,t})\subset\mathcal{M}$ be the ball of radius $\delta$ centred at $\nu_{d,t}$ and $N(\varepsilon,B_\delta(\nu_{d,t}),d)$ its covering number using balls of size $\varepsilon$. Then, as $\delta\to0$,
\[\int_0^1\sqrt{1+\log N(\delta\varepsilon,B_\delta(\nu_{d,t}),d)}\,\mathrm{d}\varepsilon=O(1).\]

(iii) There exist $\eta>0$, $C>0$, and $\kappa>1$ such that, whenever $d(\nu_{d,t},\omega)<\eta$,
\[E\{d^2(Y_t,\omega)\mid D=d\}-E\{d^2(Y_t,\nu_{d,t})\mid D=d\}\geq C d^\kappa(\nu_{d,t},\omega).\]
\end{assumption}

Assumption~\ref{asp:fm}(i) is a standard consistency condition for M-estimators, including \f means; see \citet[Chapter~3.2]{well:96}. It separates the population minimizer from alternatives outside any fixed neighbourhood and, together with convergence of the empirical objective, yields convergence of the empirical minimizer. The entropy condition in Assumption~\ref{asp:fm}(ii) controls the local complexity of the objective class near the minimizer, while the curvature condition in Assumption~\ref{asp:fm}(iii) quantifies how rapidly the population objective increases away from its minimum. In particular, $\kappa=2$ corresponds to quadratic growth and yields the parametric rate, whereas weaker local curvature may produce slower rates. Some example spaces require additional conditions, while the univariate Wasserstein rate follows directly from Hilbert-space projection arguments \citep{pete:19}; details and verification for the spaces considered here are provided in Section~S.4 of the Supplementary Material.

\begin{assumption}\label{asp:rc}
There exists a constant $c \in (0,1/2]$ such that $c \leq \operatorname{pr}(D=1)\leq 1-c$.
\end{assumption}
Assumption~\ref{asp:rc} is the standard overlap condition in the DID literature, ensuring positive population fractions in both groups.

\begin{theorem}
    \label{thm:GDD-rc}
    Under Assumptions~\ref{asp:ug}--\ref{asp:rc}, it holds for the estimate $\hat{\tau}$ of the GATT that
    \[d_{\mathcal{G}}(\hat{\tau},\tau)=O_p\{n^{-1/[2(\kappa-1)]}\}.\]
\end{theorem}

Under the conditions stated in Section~S.4 of the Supplementary Material, the example spaces in Examples~\ref{exm:fun}--\ref{exm:cd} have $\kappa=2$, yielding the parametric convergence rate $O_p(n^{-1/2})$.
For $\kappa=2$, this matches the first-order rate of Euclidean DID despite replacing arithmetic operations with their geodesic counterparts.

\begin{remark}[Repeated cross-sectional data]
The results also apply to repeated cross-sectional data. Following \citet{abad:05}, let $T_i$ indicate observation in the post-treatment period and set $Y_i=Y_{i,0}$ if $T_i=0$ and $Y_i=Y_{i,1}$ if $T_i=1$. For the pooled sample $\{(Y_i,D_i,T_i)\}_{i=1}^n$, Theorem~\ref{thm:GDD-id} applies with $\nu_{d,t}=E_\oplus\{Y_i\mid D_i=d,T_i=t\}$. The estimator in \eqref{eq:tauhat} applies with $\hat\nu_{d,t}=\argmin_{\omega\in\mathcal M}n_{d,t}^{-1}\sum_{i=1}^n1\{D_i=d,T_i=t\}d^2(Y_i,\omega)$, where $n_{d,t}=\sum_{i=1}^n1\{D_i=d,T_i=t\}$.
\end{remark}

\section{Implementation and simulations}\label{sec:sim}
\subsection{Implementation details and simulation scenarios}
Algorithm~\ref{alg:gdd} proceeds in two steps. The first computes the relevant \f means in the outcome space. For functional data in Example~\ref{exm:fun}, this is the usual $L^2$ mean; for univariate Wasserstein distributions in Example~\ref{exm:upd}, it is the mean of the quantile functions. For multivariate distributions under the Fisher--Rao metric and for compositional data, computation is performed in the square-root representation on the sphere using the R package \texttt{manifold} \citep{mull:20:5}. For networks and SPD matrices under flat geometries such as the Frobenius and Log-Euclidean metrics, the \f mean reduces to an entry-wise or log-domain mean. The second step applies the corresponding geodesic transport map; explicit constructions for all examples are given in Section~S.3 of the Supplementary Material.
\begin{algorithm}
\KwIn{data $\{(Y_{i, 0}, Y_{i, 1}, D_i)\}_{i=1}^n$.}
    \KwOut{Geodesic DID estimator $\hat\tau$ of the GATT $\tau$.}
    $(\hat{\nu}_{0, 0}, \hat{\nu}_{0, 1}, \hat{\nu}_{1, 0}, \hat{\nu}_{1, 1})\longleftarrow$ the estimated \f means for the control and treated groups during pre- and post-treatment periods:
    \begin{align*}
        \hat{\nu}_{0,t} &= \argmin_{\omega \in \mathcal{M}}{\frac{1}{n_0}}\sum_{i=1}^n (1-D_i)d^2(Y_{i,t},\omega),\\
        \hat{\nu}_{1,t} &= \argmin_{\omega \in \mathcal{M}}{\frac{1}{n_1}}\sum_{i=1}^n D_id^2(Y_{i,t},\omega),
    \end{align*}
    where $t\in\{0, 1\}$\;
    $\hat{\nu}_{1, 1}'=\Gamma_{\hat{\nu}_{0, 0}, \hat{\nu}_{0, 1}}(\hat{\nu}_{1, 0})\longleftarrow$ the new endpoint of the geodesic $\gamma_{\hat{\nu}_{0, 0}, \hat{\nu}_{0, 1}}$ after moving to $\hat{\nu}_{1, 0}$\;
    $\hat{\tau}=\gamma_{\hat{\nu}_{1, 1}', \hat{\nu}_{1, 1}}\longleftarrow$ the difference-in-differences estimator of the GATT.
    \caption{Geodesic difference-in-differences.}
\label{alg:gdd}
\end{algorithm}

The algorithm directly implements the identification formula in Theorem~\ref{thm:GDD-id}. It first estimates the four group-time \f means, then transports the control-group change from $\hat\nu_{1,0}$ to obtain the estimated counterfactual endpoint $\hat\nu_{1,1}'$. The estimated GATT is the geodesic from this counterfactual endpoint to the observed post-treatment mean $\hat\nu_{1,1}$ of the treated group.

We evaluate three settings: univariate distributions under the Wasserstein metric, bivariate distributions under the Fisher--Rao metric, and networks under the Frobenius metric (Examples~\ref{exm:upd},~\ref{exm:mpd}, and~\ref{exm:net}). In each simulation, $\operatorname{pr}(D_i=1)=0.25$.

We use sample sizes $n=50,200,1000$ and $Q=500$ Monte Carlo runs. For run $q$, estimation error is $d_{\mathcal G}(\hat\tau_q,\tau)$, with $d_{\mathcal G}$ defined in \eqref{eq:dg} and reference point $\nu_{1,1}'=\Gamma_{\nu_{0,0},\nu_{0,1}}(\nu_{1,0})$. 

\subsection{Univariate probability distributions}\label{subsec:mea}
Consider the space of univariate probability distributions described in Example~\ref{exm:upd}. For each unit $i$ and time $t\in\{0,1\}$, we generate a Gaussian distribution truncated to the interval $[-3,3]$ with mean $\mu_{i,t}$ and standard deviation $\sigma_{i,t}$. The location parameters are drawn independently as $\mu_{i,t}\sim N(\alpha_2 t,1)$, with scale parameters  $\sigma_{i,t}=\alpha_1+\beta D_i t$, so that treated units exhibit an increase in dispersion in the post-treatment period. Throughout, we set $\alpha_1=\alpha_2=\beta=1$.

As described in Section~S.3 of the Supplementary Material, the geodesic transport map for univariate distributions is induced by optimal transport. The quantile function at the starting point of $\tau=\gamma_{\nu_{1,1}',\nu_{1,1}}$ is $F_{\nu_{1,1}'}^{-1}=F_{\nu_{0,1}}^{-1}\circ F_{\nu_{0,0}}\circ F_{\nu_{1,0}}^{-1}$. Here $\nu_{0,t}$ and $\nu_{1,t}$ are truncated Gaussian distributions with common mean $\alpha_2t$ and respective standard deviations $\alpha_1$ and $\alpha_1+\beta t$. To reflect applications in which the distributional outcomes are observed only through samples, we independently draw 100 observations from every $Y_{i,t}$ and use the resulting empirical measures in the geodesic DID estimator. This introduces within-distribution approximation error in addition to the sampling error across units \citep{zhou:23}. For an absolutely continuous univariate distribution with density bounded away from zero on its support, the empirical $2$-Wasserstein error decreases at the usual parametric rate as the number of within-distribution observations increases \citep{bigo:18}.

The number of observations used to represent each distribution and the number of DID units therefore control distinct error components: the former governs empirical-measure approximation, whereas the latter governs estimation of the group-time \f means. Increasing the number of observations per distribution reduces the approximation component without changing the DID design.

\begin{figure}
	\figuresize{0.8}
	\figurebox{20pc}{}{}[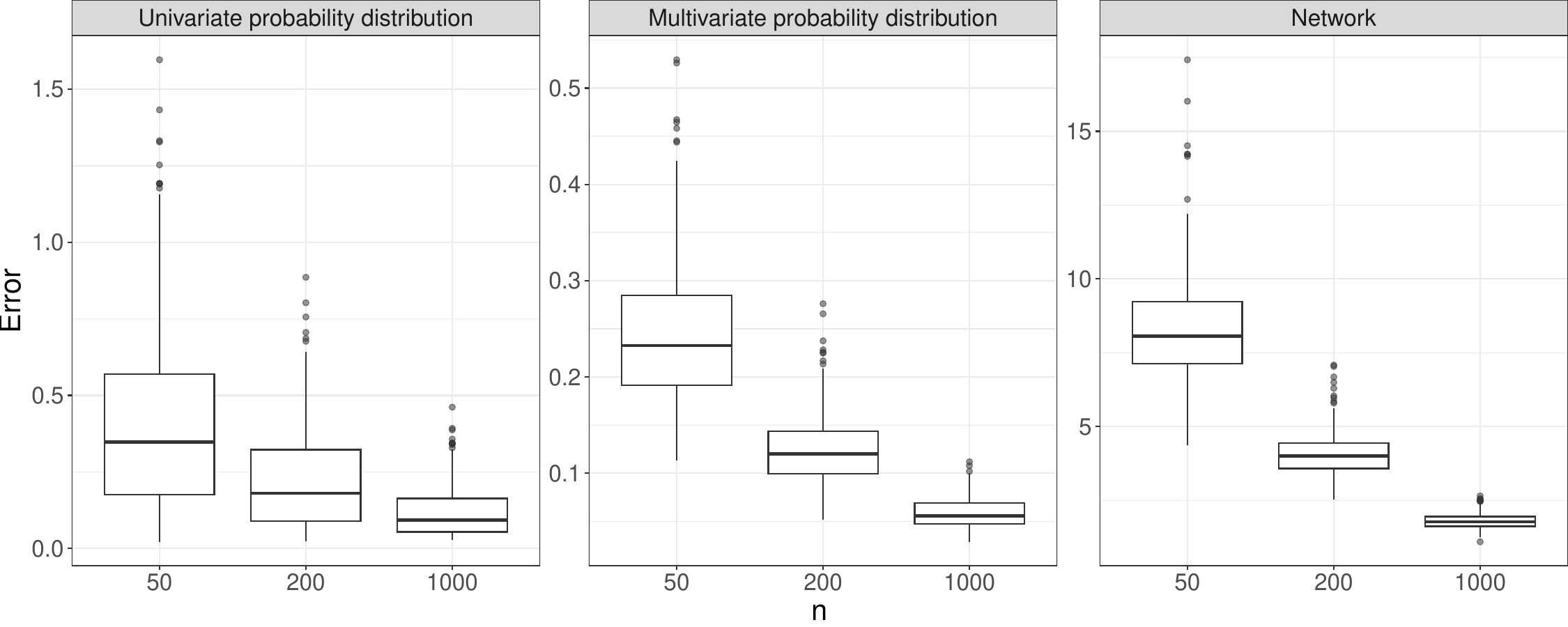]
    \caption{Boxplots of estimation errors with different sample sizes for univariate probability distributions (left), multivariate probability distributions (middle) and networks (right).}
    \figalt{Three boxplot panels show estimation errors decreasing as the sample size increases from 50 to 200 and 1000 for univariate distributions, multivariate distributions, and networks.}
    \label{fig:sim}
\end{figure}

The left panel of Figure~\ref{fig:sim} shows that the estimation error decreases as $n$ increases, demonstrating convergence of the estimated GATT towards its target despite the additional approximation error from the empirical measures. To assess the empirical convergence rate, we regress the logarithm of the average estimation error on $\log n$. The resulting slope is $-0.412$, compared with the theoretical benchmark of $-0.5$; the difference likely reflects the use of empirical measures in place of the unknown distributional outcomes.

\subsection{Multivariate probability distributions}
Consider multivariate probability distributions under the Fisher--Rao metric described in Example~\ref{exm:mpd}. For each unit $i$ and time $t\in\{0,1\}$, the outcome $Y_{i,t}$ is a bivariate Gaussian mixture distribution with density $\pi_{i,t}\phi(x;\mu_{1,i,t},\Sigma)+(1-\pi_{i,t})\phi(x;\mu_{2,i,t},\Sigma)$, where $\phi(\cdot;\mu,\Sigma)$ denotes the bivariate Gaussian density with mean $\mu$ and covariance $\Sigma=0.5I_2$. We generate $\xi_i\sim N(0,I_2)$ and set $\mu_{1,i,t}=\xi_i+\alpha_2t\,1_2$ and $\mu_{2,i,t}=\xi_i-\alpha_2t\,1_2$, where $1_2=(1,1)^\T$ and $\alpha_2=1$. Define $\pi_{i,t}=0.5+\beta D_it$, with $\beta=0.2$. Thus untreated units evolve according to a common time trend, while treated units experience a structural change in mixture weights in the post-treatment period. For each $(i,t)$, we independently sample $N=500$ observations from $Y_{i,t}$ and construct kernel density estimators. The resulting empirical densities are mapped to the positive orthant of the Hilbert sphere under square-root representation. The geodesic transport map corresponds to the parallel transport construction under the Fisher--Rao geometry; see Section~S.3 of the Supplementary Material. We use the Fisher--Rao metric here because its square-root representation yields explicit and computationally convenient geodesics in higher dimensions. This simulation provides one multivariate illustration and is not intended as a general recommendation of the Fisher--Rao metric over the $2$-Wasserstein metric.

The middle panel of Fig.~\ref{fig:sim} shows steadily decreasing estimation error as $n$ increases under the Fisher--Rao geometry. Regressing the logarithm of the average estimation error on $\log n$ gives a slope of $-0.473$, close to the theoretical benchmark of $-0.5$. The remaining difference is consistent with the additional approximation error arising from the use of kernel density estimates, analogous to the empirical-measure approximation in Section~\ref{subsec:mea}.

\subsection{Networks}
Consider the network space in Example~\ref{exm:net}, with $m\times m$ symmetric, centred graph-Laplacian outcomes \citep{mull:22:11}. We use a weighted stochastic block model with community-membership vector $z=(z_1^\T,z_2^\T)^\T$, where $z_l$ has length $m_l$ and all entries equal $l$, for $l=1,2$, with $m_1+m_2=m$. Edges occur independently with probabilities given by the corresponding entries of $P=(p_{ll'})_{l,l'=1}^2$.
Conditional on an edge being present, its weight is $\alpha_1+\alpha_2t+\alpha_3D_i+\beta D_it+\epsilon_{i,t,jk}$, where $\alpha_1=\alpha_2=\alpha_3=\beta=1$ and the $\epsilon_{i,t,jk}$ are independently drawn from the uniform distribution on $[-1,1]$.

The goal is to estimate the GATT $\tau=\gamma_{\nu_{1,1}', \nu_{1,1}}$, where $\nu_{d,t}=\Eo\{Y_{i, t}\mid D_i=d\}$ and $\nu_{1,1}'=\Gamma_{\nu_{0, 0}, \nu_{0, 1}}(\nu_{1, 0})=\nu_{1, 0}+(\nu_{0, 1}-\nu_{0, 0})$; see Section~S.3 of the Supplementary Material. Using $(\nu_{d, t})_{jk}$ to denote the $(j, k)$th entry of the graph Laplacian $\nu_{d, t}$, under the Frobenius norm,
\[(\nu_{0, t})_{jk}=-p_{ll'}(\alpha_1+\alpha_2t),\quad(\nu_{1,t})_{jk}=-p_{ll'}(\alpha_1+\alpha_2t+\alpha_3+\beta t)\]
for $1\leq j\neq k\leq m$, where $l, l'\in\{1, 2\}$ represent the community memberships of nodes $j$ and $k$, respectively. The diagonal entries of $\nu_{d, t}$ are $(\nu_{d, t})_{jj} = -\sum_{k \neq j}(\nu_{d, t})_{jk}$. 

For $m_1=m_2=5$, $p_{11}=p_{22}=0.5$, and $p_{12}=p_{21}=0.2$, the right panel of Fig.~\ref{fig:sim} shows that the estimation error decreases as the sample size increases. A least-squares regression of the logarithm of the average estimation error on $\log n$ gives a slope of $-0.509$, closely matching the theoretical value of $-0.5$. 

Across the three settings, the empirical slopes are close to the $-1/2$ benchmark. Agreement is closest for networks, whose outcomes are observed directly, whereas the distributional settings also involve empirical-measure or density-estimation error.

\section{Data applications}\label{sec:app}
\subsection{Impact of the collapse of the Soviet Union on age-at-death distributions}\label{subsec:aad}
The collapse of the Soviet Union in 1991 profoundly affected population dynamics, with dramatic declines in life expectancy observed across many former Soviet states. In Russia alone, life expectancy at birth between 1990 and 1994 dropped significantly from 63.8 to 57.6 for men  and from 74.4 to 71.0 for women \citep{leon:97}. Similar patterns were observed in other former Soviet states \citep{brai:05}. Adopting the entire age-at-death distribution as outcome provides deeper insights into trends of human longevity. To study the causal effect of the collapse, we applied the geodesic DID model with age-at-death distributions as the outcome.

The \citeauthor{HMD} provides life tables at one-, five-, or ten-year intervals
for various countries, including the six countries that were previously part of the former Soviet Union and 19 Western European countries. The six countries of the former Soviet Union form the intervention group, as they were subject to the collapse of the Soviet Union, while the 19 Western European countries serve as the control group. Further details are presented in Table~\ref{tab:ew}.

\begin{table}[H]
    \tbl{Six Eastern European countries that were part of the former Soviet Union and 19 Western European countries}
    {\begin{tabular}{cl}
        Eastern Europe & \makecell[l]{Belarus, Estonia, Latvia, Lithuania, Russia, Ukraine} \\\noalign{\vskip 5pt}
        Western Europe & \makecell[l]{Austria, Belgium, Denmark, Finland, France, Germany, Greece, Iceland, \\Ireland, Italy, Luxembourg, Netherlands, Norway, Portugal, Slovenia, \\Spain, Sweden, Switzerland, United Kingdom}
    \end{tabular}}
    \label{tab:ew}
\end{table}

\begin{figure}
  \centering
  \begin{minipage}{0.8\textwidth}
    \centering
    \figuresize{1}
	\figurebox{20pc}{}{}[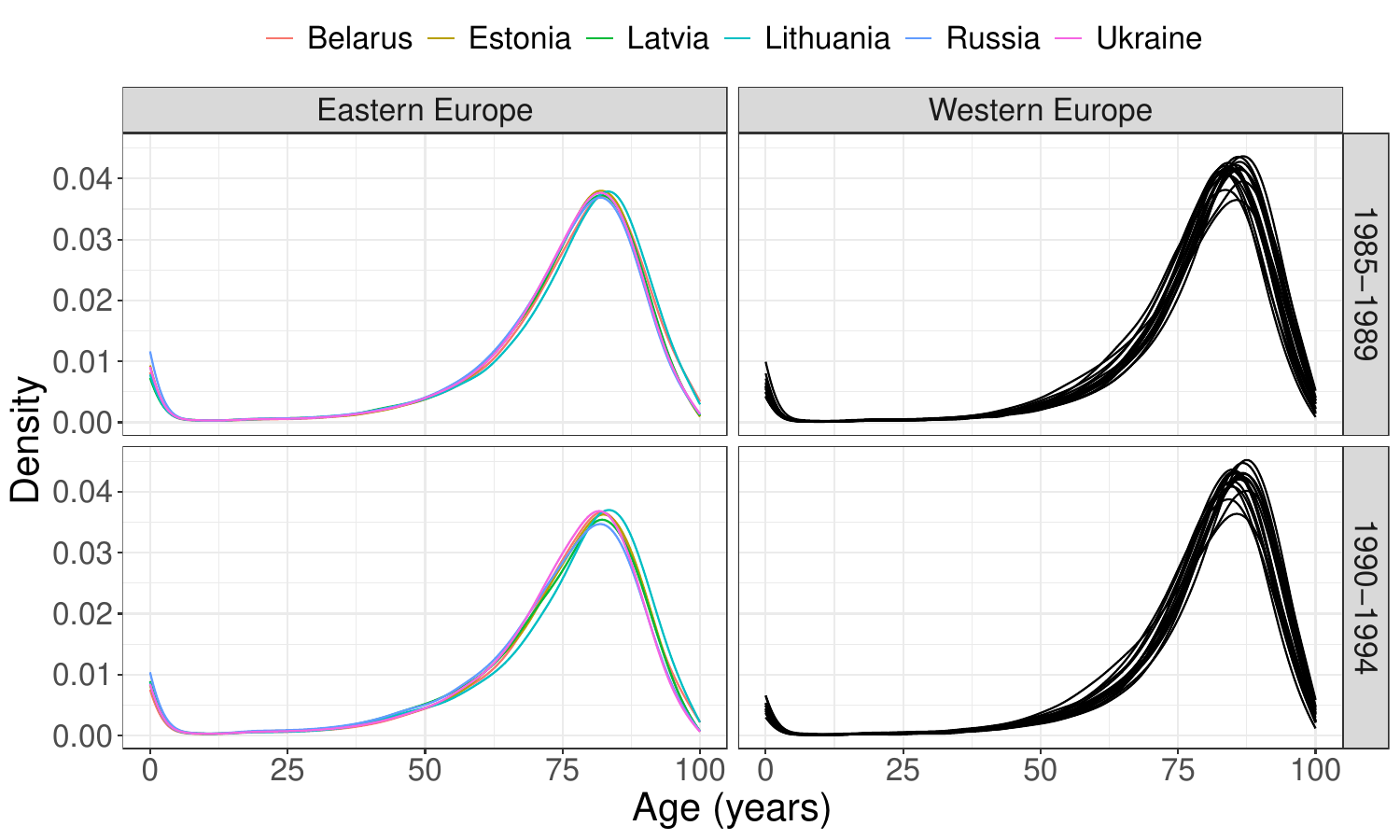]
	    (a)
  \end{minipage}\par
  \begin{minipage}{0.8\textwidth}
    \centering
    \figuresize{1}
	\figurebox{20pc}{}{}[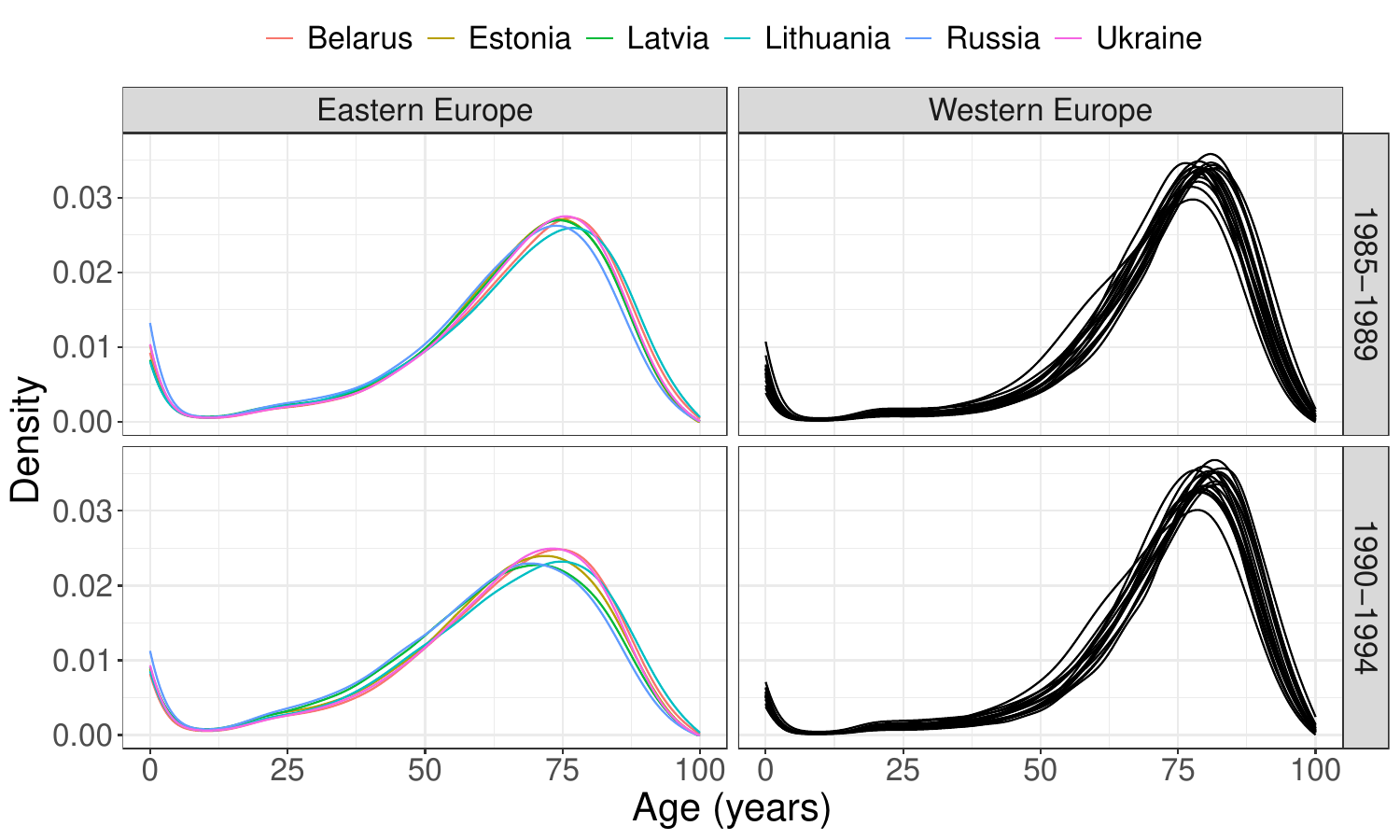]
	    (b)
  \end{minipage}
  \caption{Age-at-death distributions of (a) females and (b) males for Eastern and Western European countries before and after the collapse of the Soviet Union.}
  \figalt{Female and male density panels compare Eastern and Western European age-at-death distributions in 1985--1989 and 1990--1994. The Eastern European densities shift towards younger ages, especially for males.}
  \label{fig:d}
\end{figure}

For each country, the age-at-death distributions are derived from histograms of death counts by age, which are available in five-year intervals. These histograms are smoothed into continuous age-at-death densities using the \texttt{CreateDensity} function in the \texttt{frechet} package \citep{chen:20}. We analyse these densities for two time periods: 1985--1989 (before the collapse) and 1990--1994 (after the collapse). Figure~\ref{fig:d} illustrates the densities of the age-at-death distributions for females and males in both intervention and control groups. In the intervention group, a clear leftward shift in the densities is observed for both genders, reflecting a reduction in life expectancy after the collapse. The shift is notably more pronounced for males.

\begin{figure}
    \figuresize{0.8}
	\figurebox{20pc}{}{}[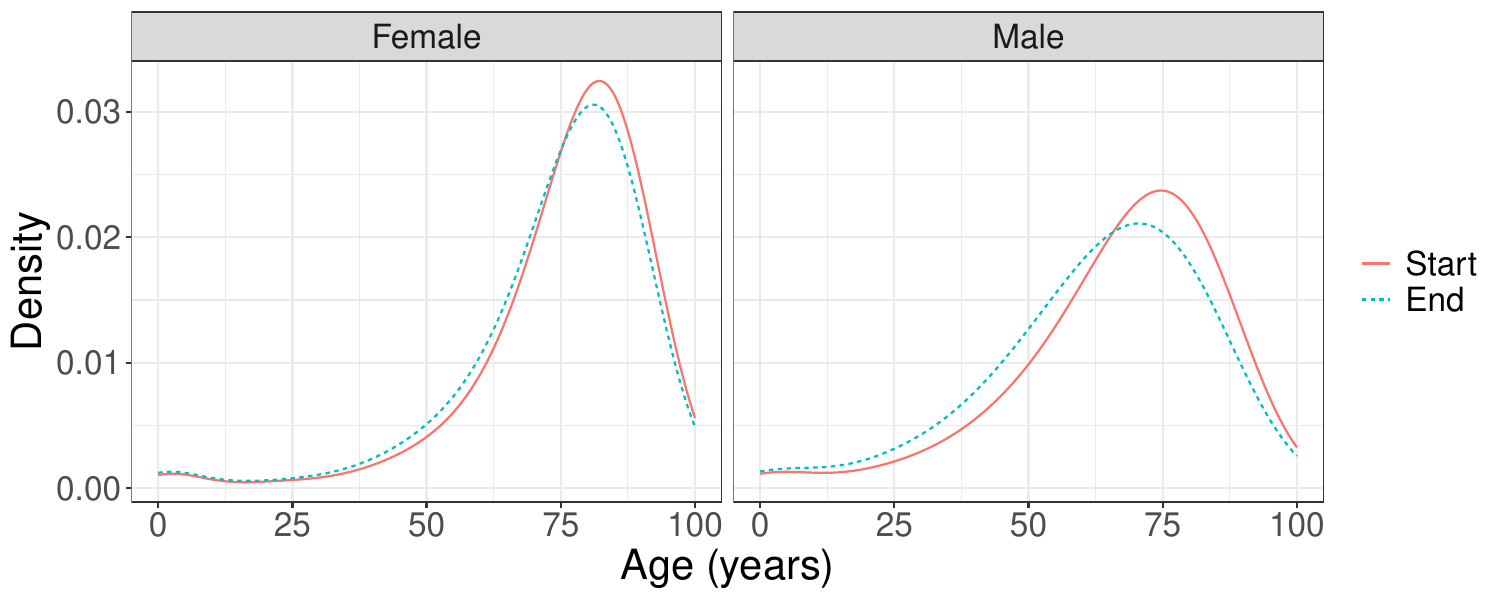]
    \caption{Estimated geodesic average treatment effect on the treated (GATT) for age-at-death distributions of females (left) and males (right). The solid red curve represents the age-at-death density at the starting point of the estimated GATT, while the dashed blue curve represents the density at the ending point.}
    \figalt{The ending density is shifted towards younger ages relative to the starting density in both panels, with a larger separation for males than for females.}
    \label{fig:dse}
\end{figure}

Using the geodesic DID model, we estimate the GATT for age-at-death distributions. The estimated GATTs for females and males are shown in Figure~\ref{fig:dse}. The results indicate a causal effect of the Soviet Union's collapse on age-at-death distributions, as evidenced by the differences between the starting and ending points of the geodesic that represents the estimate of GATT. For females, the age-at-death densities shift to the left, reflecting reduced life expectancy. For males, the effect is even stronger, which may be attributed to increased susceptibility of males to the social and economic upheavals during this period.

This application illustrates the value of distribution-valued outcomes: the estimated GATT records changes across the full mortality distribution, including its shape and location, rather than reducing the effect to a scalar change in life expectancy.

\begin{figure}[H]
    \figuresize{0.8}
		\figurebox{20pc}{}{}[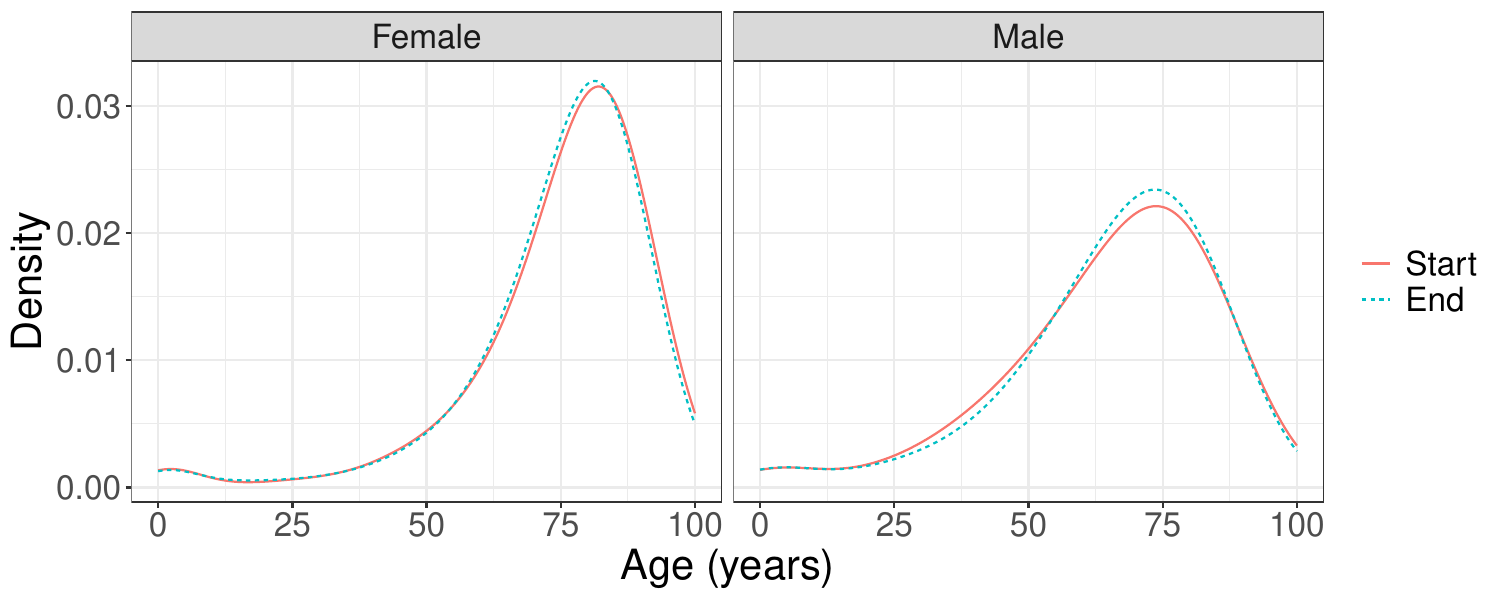]
    \caption{Pre-treatment assessment of the parallel trends assumption using age-at-death distributions from 1980--1984 and 1985--1989. The solid red curve represents the age-at-death density at the starting point of the estimated GATT, while the dashed blue curve represents the density at the ending point.}
    \figalt{The starting and ending pre-treatment GATT densities nearly overlap for females and males, with only small differences near their peaks.}
    \label{fig:dse1}
\end{figure}

To assess the plausibility of the parallel trends assumption (Assumption~\ref{asp:pt}), we estimate the GATT using two pre-treatment periods, 1980--1984 and 1985--1989, both of which precede the Soviet Union's collapse. Under parallel trends, this placebo estimand is the identity geodesic, so the distance between its starting and ending points summarizes differential evolution between the groups before treatment. Figure~\ref{fig:dse1} presents the resulting pre-treatment GATT for females (left) and males (right). The close alignment between the starting and ending densities suggests the absence of pronounced differential evolution between groups prior to treatment. Although the parallel trends assumption is inherently counterfactual and cannot be formally verified, the small magnitude of the estimated pre-treatment GATT provides empirical support for its plausibility.

\FloatBarrier
\subsection{Impact of U.S. electricity market liberalization on electricity generation}
We analyse U.S. electricity generation data, publicly available from the U.S. Energy Information Administration's website (\texttt{\url{https://www.eia.gov/electricity/data/state/}}). This dataset provides annual electricity generation figures by energy source for each state from 1990 to 2023. We consider the following three categories of energy sources: fossil, combining ``coal'' and ``petroleum''; nuclear; and renewables, combining ``hydroelectric conventional'', ``solar thermal and photovoltaic'', ``geothermal'', ``wind'', and ``wood and wood-derived fuels''. For each state and year, generation in these categories is normalized by their combined generation to form a three-dimensional compositional vector.  The component-wise square root transformation as described in Example~\ref{exm:cd} then maps these compositional outcomes onto the positive orthant of the sphere $\mathcal{S}_+^2 \subset \real^3$, which we endow with the geodesic metric.

In the early 1990s, vertically integrated monopolies controlled approximately $92\%$ of U.S. electricity generation and consumption. Under the regulatory compact, these utilities provided reliable access at cost-based prices, reflecting economies of scale in infrastructure and network externalities in electricity distribution \citep{josk:97}. The 1992 National Energy Policy Act promoted competition in generation, and Federal Energy Regulatory Commission Order 888 in 1996 prevented utilities from favouring their own generators through control of transmission infrastructure, thereby lowering barriers to entry for third-party producers. States nevertheless retained authority over whether to adopt market liberalization \citep{josk:97}. Between 1996 and 2000, 23 states adopted liberalization and form the treated group; the remaining 27 states serve as controls (Table~\ref{tab:eml}).

\begin{table}[H]
    \tbl{23 treated and 27 control states}
    {\begin{tabular}{cl}
        Treated states & \makecell[l]{Arizona, Arkansas, California, Connecticut, Delaware, Illinois, Maine, \\Maryland, Massachusetts, Montana, Nevada, New Hampshire, New Jersey, \\New Mexico, New York, Ohio, Oklahoma, Oregon, Pennsylvania, \\Rhode Island, Texas, Virginia, West Virginia} \\\noalign{\vskip 5pt}
        Control states & \makecell[l]{Alabama, Alaska, Colorado, Florida, Georgia, Hawaii, Idaho, Indiana, Iowa, \\Kansas, Kentucky, Louisiana, Michigan, Minnesota, Mississippi, Missouri, \\Nebraska, North Carolina, North Dakota, South Carolina, South Dakota, \\Tennessee, Utah, Vermont, Washington, Wisconsin, Wyoming}
    \end{tabular}}
    \label{tab:eml}
\end{table}

\begin{figure}
  \centering
  \begin{minipage}{0.48\textwidth}
    \centering
    \figuresize{1}
	\figurebox{20pc}{}{}[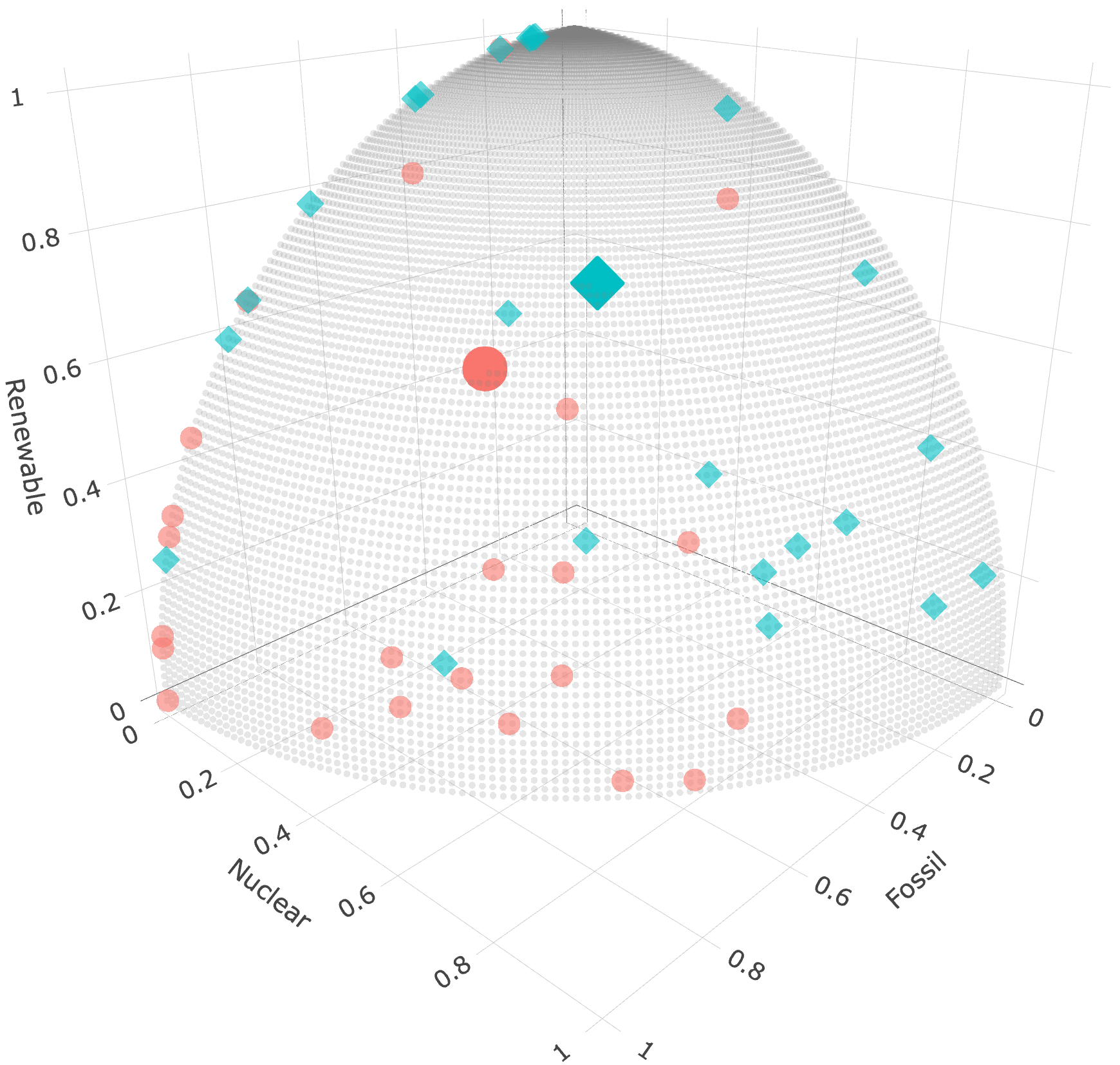]
	    (a)
  \end{minipage}
  \hfill
  \begin{minipage}{0.48\textwidth}
    \centering
    \figuresize{1}
	\figurebox{20pc}{}{}[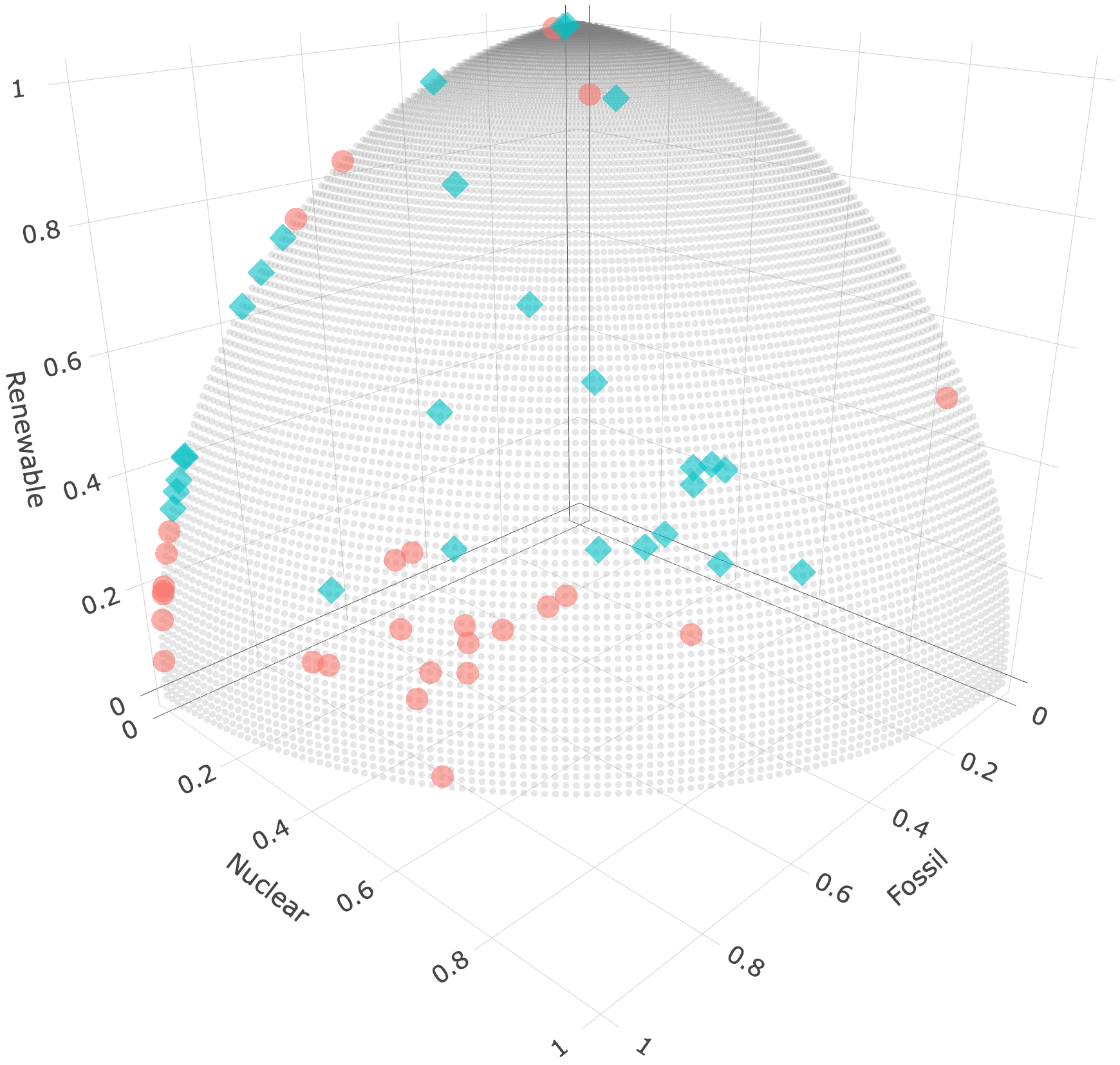]
	    (b)
  \end{minipage}
  \caption{Compositions of energy sources for electricity generation for (a) treated and (b) control states. Circles and diamonds represent compositions before and after the electricity market liberalization, respectively. The two larger markers correspond to the starting and ending points of the estimated GATT.}
  \figalt{Sphere plots compare 1995 circles with 2020 diamonds for treated and control states. The larger treated-group markers show the estimated GATT shifting away from fossil generation and towards renewables.}
  \label{fig:com1}
\end{figure}

\begin{figure}
    \figuresize{1}
	\figurebox{20pc}{}{}[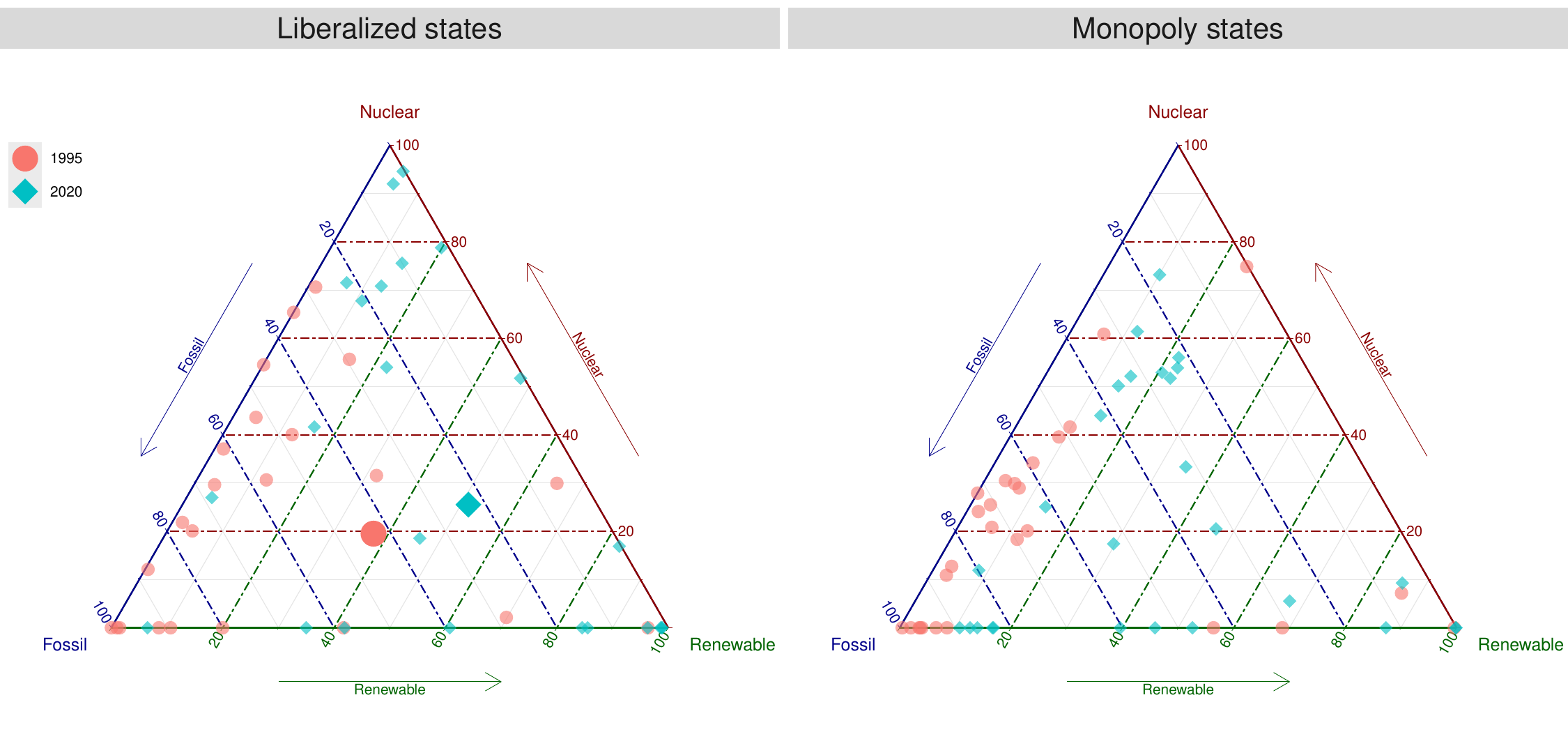]
    \caption{Ternary plots illustrating compositions of energy sources for electricity generation for treated (left) and control (right) states. Circles and diamonds represent compositions before and after the electricity market liberalization, respectively. The two larger markers correspond to the starting and ending points of the estimated GATT.}
    \figalt{Ternary plots compare fossil, nuclear, and renewable shares in 1995 and 2020. The larger treated-group markers indicate lower fossil and higher renewable shares at the end of the estimated GATT.}
    \label{fig:com2}
\end{figure}

We assess the effect of electricity market liberalization on generation composition, using 1995 as the pre-treatment period and 2020 as the post-treatment period. The 25-year interval allows time for gradual effects arising from investment, generation-fleet turnover, and changes in fuel mix \citep{lee:20:1}. Figures~\ref{fig:com1} and~\ref{fig:com2} show the sphere and ternary representations. Treated states exhibit lower fossil generation and higher nuclear and renewable generation after liberalization.

Because 1995 precedes adoption in every treated state and all treated states had adopted by 2000, the 1995 and 2020 observations provide common pre- and post-treatment periods across the treated group.

Using the geodesic DID model, we estimate the GATT for these compositional outcomes. The estimate is visualized in the left panels of Figs.~\ref{fig:com1} and~\ref{fig:com2}. The starting and ending compositions of the estimated GATT are $(0.432, 0.195, 0.373)^\T$ and $(0.231, 0.255, 0.514)^\T$, respectively. Relative to the transported counterfactual composition, the observed post-liberalization composition has a 20.1 percentage-point lower fossil share, a 6.0 percentage-point higher nuclear share, and a 14.1 percentage-point higher renewable share. The results demonstrate a causal effect of electricity market liberalization on the composition of energy sources for electricity generation. Specifically, renewable energy sources gained a larger share following market liberalization, while the share of fossil fuels declined. These findings suggest that deregulation incentivized cleaner energy practices and supported a transition toward more sustainable energy systems.

To examine the plausibility of the parallel trends assumption, we compare compositional outcomes for the periods 1990 and 1995, with 1990 being the earliest available year in the dataset. The estimated GATT between these two pre-treatment periods has starting composition $(0.694, 0.181, 0.125)^\T$ and ending composition $(0.685, 0.179, 0.136)^\T$. The relatively small magnitude of these changes suggests the absence of pronounced differential pre-treatment trends and provides empirical support for the reasonableness of the parallel trends assumption. The corresponding spherical distances are $0.216$ post-treatment and $0.016$ in the pre-treatment comparison, showing that the estimated post-treatment change is substantially larger.

\section{Concluding remarks}\label{sec:dis}

For outcome spaces or metrics beyond those considered here, Assumptions~\ref{asp:iso} and~\ref{asp:com} must be verified case by case. Consequently, the convergence guarantee in Theorem~\ref{thm:GDD-rc} does not automatically extend to every metric on a given outcome class.

Several challenges remain. Inference for the causal estimand in standard DID has been well-studied under the assumption of independent sampling and extended to allow for serial correlation using clustering methods \citep{bert:04}. However, these approaches cannot be directly applied to non-Euclidean outcomes due to the lack of linear structure. Further research will be needed to address these and related issues.

Relaxing the parallel trends assumption, which has been explored for Euclidean settings \citep{abad:05}, is another promising future direction. In non-Euclidean spaces, parallel trends violations could be accommodated by extending the geodesic DID framework to incorporate model-based adjustments or alternative causal assumptions. This extension could enhance its applicability for scenarios where the parallel trends assumption does not hold. An extension to metric spaces that do not support geodesics and may not be length spaces is another theoretical challenge for future exploration. 

\section*{Acknowledgement}
We wish to thank the editor and the reviewers for their valuable comments and suggestions, which have significantly improved the quality of the paper. D.K. was supported in part by a Japan Society for the Promotion of Science KAKENHI grant and H.G.M. by a U.S. National Science Foundation grant. 

\section*{Supplementary material}
\label{SM}
The Supplementary Material includes the extension to staggered difference-in-differences settings, a discussion of phylogenetic tree space, explicit constructions of the corresponding geodesic transport maps, a discussion and verification of the model assumptions, and all technical proofs.

\single
\bibliographystyle{rss}
\bibliography{collection}
\double

\clearpage
\begin{center}
{\large\bfseries SUPPLEMENTARY MATERIAL}
\end{center}

\setcounter{section}{0}
\renewcommand{\thesection}{S.\arabic{section}}
\renewcommand{\thesubsection}{S.\arabic{section}.\arabic{subsection}}
\renewcommand{\theHsection}{S.\arabic{section}}
\renewcommand{\theHsubsection}{S.\arabic{section}.\arabic{subsection}}

\setcounter{assumption}{0}
\setcounter{theorem}{0}
\setcounter{corollary}{0}
\setcounter{proposition}{0}
\setcounter{definition}{0}
\setcounter{remark}{0}
\setcounter{example}{0}
\setcounter{figure}{0}
\setcounter{table}{0}
\setcounter{equation}{0}
\renewcommand{\theassumption}{S\arabic{assumption}}
\renewcommand{\thetheorem}{S\arabic{theorem}}
\renewcommand{\thecorollary}{S\arabic{corollary}}
\renewcommand{\theproposition}{S\arabic{proposition}}
\renewcommand{\thedefinition}{S\arabic{definition}}
\renewcommand{\theremark}{S\arabic{remark}}
\renewcommand{\theexample}{S\arabic{example}}
\renewcommand{\thefigure}{S\arabic{figure}}
\renewcommand{\thetable}{S\arabic{table}}
\renewcommand{\theequation}{S\arabic{equation}}
\renewcommand{\theHassumption}{S\arabic{assumption}}
\renewcommand{\theHtheorem}{S\arabic{theorem}}
\renewcommand{\theHcorollary}{S\arabic{corollary}}
\renewcommand{\theHproposition}{S\arabic{proposition}}
\renewcommand{\theHdefinition}{S\arabic{definition}}
\renewcommand{\theHremark}{S\arabic{remark}}
\renewcommand{\theHexample}{S\arabic{example}}
\renewcommand{\theHfigure}{S\arabic{figure}}
\renewcommand{\theHtable}{S\arabic{table}}
\renewcommand{\theHequation}{S\arabic{equation}}

\section{Extension to the staggered case}\label{sec:ext}
\subsection{Staggered geodesic DID}
The staggered case features multiple time periods and variations in treatment timing. The theoretical results provided in this section can be seen as a natural extension of those in \citet{CaSa21} for Euclidean data to the case of outcomes in uniquely geodesic spaces. We introduce notation similar to that in \citet{CaSa21}. Let $D_{i,t}$ be a binary variable equal to one if unit $i \in \{1,\ldots,n\}$ is treated in period $t \in \{0,1,\ldots, T\}$ and equal to zero otherwise, where we require
\begin{assumption}[Irreversibility of treatment]\label{ass:irrev-treat}
For any unit $i=1,\ldots,n$, (i) $D_{i,0}=0$; and (ii) if $D_{i,t-1}=1$, then $D_{i,t}=1$ for $t=1,\ldots,T$.
\end{assumption}
Assumption~\ref{ass:irrev-treat} means that every unit is untreated at time $t=0$ and that once a unit becomes treated, the unit will remain treated in the following periods. This assumption is commonly referred to as staggered treatment adoption.

Let $G_i$ denote the time period when unit $i$ first becomes treated. If a unit does not become treated in any time period, we set $G_i = \infty$. For all units that are treated, $G_i$ indicates which group unit $i$ belongs to. Let $G_{i,g}=1\{G_i=g\}$ indicate that unit $i$ is first treated in period $g$, and let $C_i=1\{G_i=\infty\}=1-D_{i,T}$ indicate that unit $i$ is never treated.

For the observed outcome $Y_{i,t} \in \mathcal{M}$ for each unit $i$ at time $t$, we assume
\begin{align*}
Y_{i,t} &=
\begin{cases}
Y_{i,t}(0) & \text{if}\ C_i=1,\\
Y_{i,t}(g) & \text{if}\ C_i=0,\ G_{i,g}=1,
\end{cases}
\end{align*}
where $Y_{i,t}(0)$ denotes unit $i$'s untreated potential outcome at time $t$ if the unit remains untreated through time period $T$, i.e., if it is never treated across all available time periods, and $Y_{i,t}(g)$ denotes the potential outcome for unit $i$ at time $t$ if the unit were first treated in time period $g=1,\ldots,T$. Assume that $\{(Y_{i,0}, Y_{i,1}, \ldots, Y_{i,T}, D_{i,0},D_{i,1},\ldots, D_{i,T})\}_{i=1}^n$ is an i.i.d. sample drawn from a joint distribution $P$, where a generic random variable distributed according to $P$ is $(Y_0, Y_1, \ldots, Y_T, D_0, D_1, \ldots, D_T)$, where $Y_t \in \mathcal{M}$ and $D_t \in \{0, 1\}$, $t=0,\ldots, T$.
We now define causal effects under staggered treatment adoption.

\begin{definition}[Group-time GATT]
    The geodesic average treatment effect for units who are members of a group $g$ at a time period $t$ is defined as the geodesic connecting $\Eo\{Y_{i,t}(0)\mid G_{i,g}=1\}$ and $\Eo\{Y_{i,t}(g)\mid G_{i,g}=1\}$, i.e.,
    $\tau(g,t)=\gamma_{\Eo\{Y_{i,t}(0)\mid G_{i,g}=1\}, \Eo\{Y_{i,t}(g)\mid G_{i,g}=1\}}$.
    We refer to  $\tau(g,t)$ as the group-time GATT of a group $g$ at a time period $t$.
\end{definition}

\subsection{Identification}
In this subsection, we provide results for the identification of the group-time GATT $\tau(g,t)$. Let $\widetilde{\mathcal{G}}$ denote the support of $\{G_i\}_{i=1}^n$. We set $\bar{g}=\max_{1 \leq i \leq n}G_i$ if $G_i \leq T$ for all $i$ and $\bar{g}=\infty$ if $G_i=\infty$ for some $i$. Define $\mathcal{G} = \widetilde{\mathcal{G}}\backslash \{\bar{g}\} \subset \{1,2,\ldots,T\}$. We require
\begin{assumption}[Limited treatment anticipation]\label{ass:lim-t-anti}
There is a known $\delta \geq 0$ such that
\[
\Eo\{Y_{i,t}(g)\mid G_{i,g}=1\} = \Eo\{Y_{i,t}(0)\mid G_{i,g}=1\}
\]
for all $g \in \mathcal{G}$, $t\in \{0,1,\ldots,T\}$ with $t < g-\delta$.
\end{assumption}

\begin{assumption}[Parallel trends based on a never-treated group]\label{ass:pt-nev}
    Let $\delta$ be as defined in Assumption~\ref{ass:lim-t-anti}. For each $g \in \mathcal{G}$ and $t \in \{1,\ldots,T\}$ such that $t \geq g-\delta$,
    \begin{align*}
    \Gamma_{\alpha, \beta}(\Eo\{Y_{i,t-1}(0)\mid G_{i,g}=1\}) &= \Eo\{Y_{i,t}(0)\mid G_{i,g}=1\},
    \end{align*}
    where $\alpha = \Eo\{Y_{i,t-1}(0)\mid C_i=1\}$ and $\beta = \Eo\{Y_{i,t}(0)\mid C_i=1\}$.
\end{assumption}

\begin{assumption}[Parallel trends based on a not-yet-treated group]\label{ass:pt-ny}
    Let $\delta$ be as defined in Assumption~\ref{ass:lim-t-anti}. For each $g \in \mathcal{G}$ and each $(s,t) \in \{1,\ldots,T\} \times \{1,\ldots,T\}$ such that $t \geq g-\delta$ and $t+\delta \leq s < \bar{g}$,
    \begin{align*}
    \Gamma_{\alpha, \beta}(\Eo\{Y_{i,t-1}(0)\mid G_{i,g}=1\}) &= \Eo\{Y_{i,t}(0)\mid G_{i,g}=1\},
    \end{align*}
    where $\alpha = \Eo\{Y_{i,t-1}(0)\mid D_{i,s}=0,G_{i,g}=0\}$ and $\beta = \Eo\{Y_{i,t}(0)\mid D_{i,s}=0,G_{i,g}=0\}$.
\end{assumption}

Assumption~\ref{ass:lim-t-anti} restricts anticipation of the treatment for all treated groups. When $\delta = 0$, it corresponds to a ``no-anticipation'' assumption. Assumptions~\ref{ass:pt-nev} and~\ref{ass:pt-ny} restrict the evolution of untreated potential outcomes and are generalizations of the parallel trends assumption (Assumption~2 in the main paper). Assumption~\ref{ass:pt-nev} means that the average outcomes for the group first treated in period $g$ and for the never-treated group have parallel paths in the absence of treatment. Assumption~\ref{ass:pt-ny} imposes parallel trends between group $g$ and not-yet-treated groups by time $t+\delta$. For further discussion of Assumptions~\ref{ass:pt-nev} and~\ref{ass:pt-ny} for the Euclidean special case, we refer to \citet{CaSa21}.

\begin{figure}
    \figuresize{0.6}
	\figurebox{20pc}{}{}[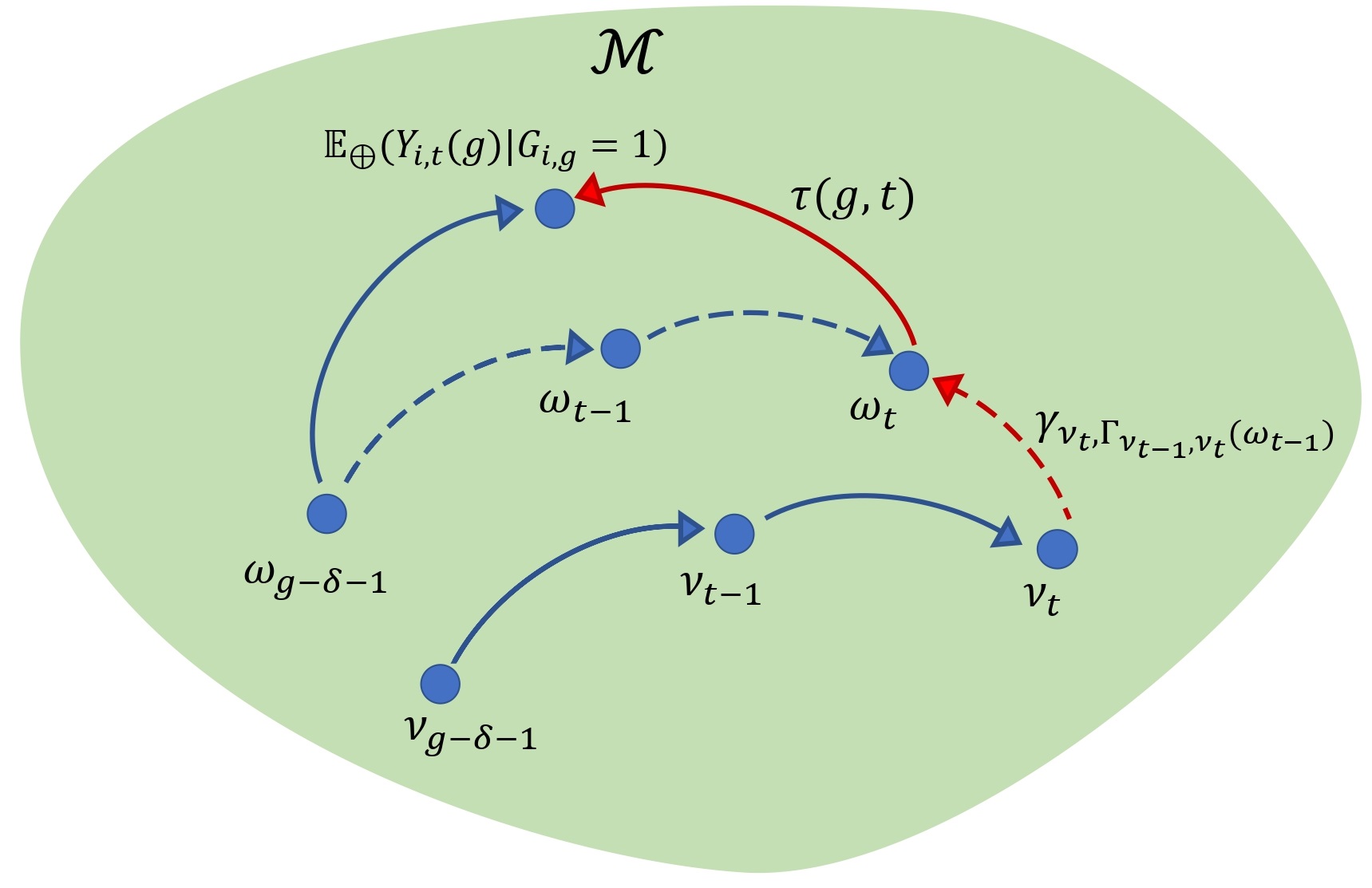]
    \caption{Illustration of staggered geodesic DID. The blue solid circles symbolize objects in the uniquely geodesic space $\mathcal{M}$. The arrows emanating from the circles represent the geodesic paths connecting them. The red arrows represent the geodesic transport from $\nu_t$ to $\omega_t$ and the proposed group-time GATT. Here, $\omega_s = \Eo\{Y_{i,s}(0)\mid G_{i,g}=1\}$. Under Assumption~\ref{ass:pt-nev}, we set $\nu_s=\Eo\{Y_{i,s}(0)\mid C_i=1\}$. Under Assumption~\ref{ass:pt-ny}, we set $\nu_s = \Eo\{Y_{i,s}(0)\mid D_{i,t+\delta}=0, G_{i,g}=0\}$.}\figalt{Diagram in $\mathcal M$ showing the paths from $\nu_{g-\delta-1}$ through $\nu_{t-1}$ to $\nu_t$, transport of the increment from $\nu_{t-1}$--$\nu_t$ to $\omega_{t-1}$--$\omega_t$, and the group-time GATT from $\omega_t$ to $E_\oplus\{Y_{i,t}(g)\mid G_{i,g}=1\}$.}
    \label{fig:GSDD-image}
\end{figure}

For a generic $\delta \geq 0$, define $\mathcal{G}_\delta = \mathcal{G} \cap \{1+\delta,2+\delta,\ldots, T+\delta\}$. As we will see in Theorem~\ref{thm:GDD-stag-id}, we can identify the group-time GATT $\tau(g,t)$. Figure~\ref{fig:GSDD-image} illustrates the identification of the group-time GATT using geodesic transport maps.

\begin{theorem}\label{thm:GDD-stag-id}
Suppose that Assumption~1 in the main paper and Assumptions~\ref{ass:irrev-treat} and~\ref{ass:lim-t-anti} hold.

(i) If Assumption~\ref{ass:pt-nev} holds, then for all $g$ and $t$ such that $g\in\mathcal{G}_\delta$, $t\in\{1,\ldots,T-\delta\}$, and $t\geq g-\delta$,
$\tau(g,t) = \gamma_{\beta_t, \Eo\{Y_{i,t}\mid G_{i,g}=1\}}$, where $\beta_t$ is defined recursively as
\begin{align*}
\beta_s &=
\begin{cases}
\Eo\{Y_{i,g-\delta-1}\mid G_{i,g}=1\} & s=g-\delta-1,\\
\Gamma_{\Eo\{Y_{i,s-1}\mid C_i=1\},\Eo\{Y_{i,s}\mid C_i=1\}}(\beta_{s-1}) & s= g-\delta,\ldots,t.
\end{cases}
\end{align*}
(ii) If Assumption~\ref{ass:pt-ny} holds, then for all $g$ and $t$ such that $g\in\mathcal{G}_\delta$, $t\in\{1,\ldots,T-\delta\}$, and $g-\delta\leq t<\bar g-\delta$,
$\tau(g,t) = \gamma_{\beta_t, \Eo\{Y_{i,t}\mid G_{i,g}=1\}}$,
where $\beta_t$ is defined recursively as
\begin{align*}
\beta_s &=
\begin{cases}
\Eo\{Y_{i,g-\delta-1}\mid G_{i,g}=1\} & s=g-\delta-1,\\
\Gamma_{\nu_{s-1},\nu_s}(\beta_{s-1}) & s=g-\delta,\ldots,t,
\end{cases}
\end{align*}
with $\nu_s = \Eo\{Y_{i,s}\mid D_{i,t+\delta}=0,G_{i,g}=0\}$.
\end{theorem}

The identification results in Theorem~\ref{thm:GDD-stag-id} become more concise if we additionally assume the following condition:
\begin{assumption}\label{ass:gt}
    For any $\alpha,\beta, \zeta, \omega \in \mathcal{M}$ with $\alpha \neq \beta$,
    \[(\Gamma_{\zeta, \beta} \circ \Gamma_{\alpha,\zeta})(\omega)= \Gamma_{\zeta, \beta}(\Gamma_{\alpha,\zeta}(\omega)) = \Gamma_{\alpha,\beta}(\omega).\]
\end{assumption}

Assumption~\ref{ass:gt} imposes a consistency condition on the geodesic transport map, ensuring that transporting via an intermediate point \(\zeta\) is equivalent to direct transport from \(\alpha\) to \(\beta\). This path-independence simplifies the analysis of geodesic operations within the DID framework. The assumption naturally holds in spaces such as Euclidean spaces, Hilbert spaces, the Wasserstein space, the space of networks, and the space of SPD matrices.

\begin{corollary}\label{cor:GDD-stag-id}
Suppose that Assumption~1 in the main paper and Assumptions~\ref{ass:irrev-treat},~\ref{ass:lim-t-anti}, and~\ref{ass:gt} hold.

(i) If Assumption~\ref{ass:pt-nev} holds, then for all $g$ and $t$ such that $g\in\mathcal{G}_\delta$, $t\in\{1,\ldots,T-\delta\}$, and $t\geq g-\delta$,
\begin{align*}
\tau(g,t)= \ominus \gamma_{\Eo\{Y_{i,g-\delta-1}\mid C_i=1\}, \Eo\{Y_{i,t}\mid C_i=1\}}
\, \oplus \gamma_{\Eo\{Y_{i,g-\delta-1}\mid G_{i,g}=1\}, \Eo\{Y_{i,t}\mid G_{i,g}=1\}}.
\end{align*}
(ii) If Assumption~\ref{ass:pt-ny} holds, then for all $g$ and $t$ such that $g\in\mathcal{G}_\delta$, $t\in\{1,\ldots,T-\delta\}$, and $g-\delta\leq t<\bar g-\delta$,
\begin{align*}
\hspace{-.4cm} \tau(g,t) = \ominus \gamma_{\Eo\{Y_{i,g-\delta-1}\mid D_{i,t+\delta}=0,G_{i,g}=0\}, \Eo\{Y_{i,t}\mid D_{i,t+\delta}=0,G_{i,g}=0\}}
 \oplus \gamma_{\Eo\{Y_{i,g-\delta-1}\mid G_{i,g}=1\}, \Eo\{Y_{i,t}\mid G_{i,g}=1\}}.
\end{align*}
\end{corollary}

\subsection{Estimation}
In this subsection, we discuss the estimation of the group-time GATT $\tau(g,t)$ under the parallel trends assumption for either a never-treated group (Assumption~\ref{ass:pt-nev}) or a not-yet-treated group (Assumption~\ref{ass:pt-ny}). The asymptotic properties of the estimators are not elaborated here, but can be established along the lines of Theorem~2 in the main paper. We first consider the estimation of $\tau(g,t)$ under Assumption~\ref{ass:pt-nev}. In this case, we can estimate $\tau(g,t)$ as $\hat{\tau}(g,t) = \gamma_{\hat{\beta}_t,\hat{\nu}_{1,t}}$ where
\begin{align*}
\hat{\beta}_s &=
\begin{cases}
\hat{\nu}_{1,g-\delta-1} & s=g-\delta-1,\\
\Gamma_{\hat{\nu}_{0,s-1},\hat{\nu}_{0,s}}(\hat{\beta}_{s-1}) & s= g-\delta,\ldots,t,
\end{cases}
\end{align*}
\begin{align*}
\hat{\nu}_{0,s} &= \argmin_{\omega \in \mathcal{M}}{\frac{1}{n_{\mathrm{nev}}}}\sum_{i=1}^n C_i\,d^2(Y_{i,s},\omega),\ n_{\mathrm{nev}}=\sum_{i=1}^n C_i,\\
\hat{\nu}_{1,s} &= \argmin_{\omega \in \mathcal{M}}{\frac{1}{n_g}}\sum_{i=1}^n G_{i,g}\,d^2(Y_{i,s},\omega),\ n_g=\sum_{i=1}^n G_{i,g}.
\end{align*}
If Assumption~\ref{ass:gt} additionally holds, then the estimator has the simpler form $\hat{\tau}(g,t) = \ominus \gamma_{\hat{\nu}_{0,g-\delta-1}, \hat{\nu}_{0,t}} \oplus \gamma_{\hat{\nu}_{1,g-\delta-1},\hat{\nu}_{1,t}}$.

Next, we consider the estimation of $\tau(g,t)$ under Assumption~\ref{ass:pt-ny}. In this case, we can estimate $\tau(g,t)$ as $\hat{\tau}(g,t) = \gamma_{\hat{\beta}_t,\hat{\nu}_{1,t}}$ where
\begin{align*}
\hat{\beta}_s &=
\begin{cases}
\hat{\nu}_{1,g-\delta-1} & s=g-\delta-1,\\
\Gamma_{\hat{\nu}_{0,s-1},\hat{\nu}_{0,s}}(\hat{\beta}_{s-1}) & s=g-\delta,\ldots,t,
\end{cases}
\end{align*}
\begin{align*}
\hat{\nu}_{0,s} &= \argmin_{\omega \in \mathcal{M}}{\frac{1}{n_{{\mathrm{ny}}}}}\sum_{i=1}^n (1-D_{i,t+\delta})(1-G_{i,g})\,d^2(Y_{i,s},\omega),\\
n_{\mathrm{ny}} &=\sum_{i=1}^n (1-D_{i,t+\delta})(1-G_{i,g}),\\
\hat{\nu}_{1,s} &= \argmin_{\omega \in \mathcal{M}}{\frac{1}{n_g}}\sum_{i=1}^n G_{i,g}\,d^2(Y_{i,s},\omega).
\end{align*}
If Assumption~\ref{ass:gt} additionally holds, then again the estimator has the simpler form
$\hat{\tau}(g,t) = \ominus \gamma_{\hat{\nu}_{0,g-\delta-1}, \hat{\nu}_{0,t}} \oplus \gamma_{\hat{\nu}_{1,g-\delta-1},\hat{\nu}_{1,t}}$.

\section{Phylogenetic trees and BHV tree space}
Phylogenetic trees are important data objects in evolutionary biology and phylogenomics, representing ancestral relationships among a fixed set of labelled taxa \citep{kapl:20}. Modern genomic studies often produce collections of trees, and the statistical analysis of such samples requires a geometric framework for tree-valued data \citep{bard:13, nye:17}.

We fix a leaf set as $\{0,1,\ldots,n\}$. The Billera–Holmes–Vogtmann (BHV) tree space $\mathcal{T}_n$ \citep{bill:01} consists of all unrooted phylogenetic trees with these labelled leaves and nonnegative edge lengths. Each tree is characterized by a combinatorial topology together with branch lengths assigned to its edges.
A convenient representation of $\mathcal{T}_n$ is obtained by embedding trees into $\mathbb{R}^N$, where $N = 2^{n} - 1$ equals the number of possible splits (bipartitions) of the leaf set. Each coordinate corresponds to a split, and a tree is represented by a vector whose coordinate equals the length of the corresponding edge if that split is present in the tree, and zero otherwise. Not all nonnegative vectors in $\mathbb{R}^N$ correspond to trees, since incompatible splits cannot occur simultaneously.

Trees sharing the same topology correspond to points in a common Euclidean orthant. If a topology contains $d$ splits, the corresponding set of trees forms a $d$-dimensional Euclidean orthant $\mathbb{R}_+^d$. Binary trees, in which all interior vertices have degree three, correspond to top-dimensional orthants, while unresolved trees lie on lower-dimensional faces. Distinct orthants meet along shared faces corresponding to trees with fewer splits, and transitions between orthants correspond combinatorially to nearest-neighbour interchange moves.

The BHV metric is defined as the intrinsic length metric induced by this orthant structure. If two trees lie in the same orthant (i.e., share the same topology), their distance is the Euclidean distance between their coordinate vectors. If they have different topologies, the distance is defined as the length of the shortest path within $\mathcal{T}_n$, where paths are piecewise Euclidean and may pass through lower-dimensional faces. Such shortest paths are geodesics. An efficient polynomial-time algorithm for computing these geodesic distances was developed by \citet{owen:11}. A basic geometric property of $\mathcal{T}_n$ is that it is globally non-positively curved and is a CAT(0) space \citep{bill:01}. As a consequence, between any two trees there exists a unique geodesic, and metric projections onto closed convex subsets are well defined. In particular, \f means of tree-valued data are uniquely defined.

Consider the BHV tree space $\mathcal{T}_3$ consisting of unrooted phylogenetic trees with four labelled leaves and nonnegative edge lengths. In this case, there are three possible binary tree topologies, each corresponding to a distinct interior edge (split). Each topology forms a one-dimensional Euclidean orthant $\mathbb{R}_+$ parameterized by the length of the interior edge. These three rays meet at a common origin corresponding to the unresolved star tree. Thus, $\mathcal{T}_3$ is isometric to a 3-spider space, that is, the union of three half-lines joined at a single point. The metric is the intrinsic length metric: if two trees lie on the same ray, their distance is the absolute difference of their edge lengths; if they lie on different rays, the geodesic passes through the origin and the distance equals the sum of their edge lengths; see also \citet{zhu:23}.

Because $\mathcal{T}_3$ is a one-dimensional CAT(0) space, between any two trees there exists a unique geodesic. Moreover, geodesics admit unique extension along each ray. Let $T_1,T_2\in\mathcal{T}_3$ and denote by $v$ the signed displacement along the ray determined by the geodesic from $T_1$ to $T_2$. For any third tree $T_3$, we define the geodesic transport map by moving $T_3$ along its ray by the same signed displacement $v$, with the convention that if the displacement crosses the origin, the path continues along the appropriate outgoing ray. Formally, if $\ell(T)$ denotes the signed coordinate of $T$ relative to the origin along its ray, then $\Gamma_{T_1,T_2}(T_3)$ is the unique tree satisfying $\ell(\Gamma_{T_1,T_2}(T_3))=\ell(T_3)+\left(\ell(T_2)-\ell(T_1)\right)$, where the sign convention accounts for transitions through the origin. This construction ensures that $\Gamma_{T_1,T_2}(T_1)=T_2$ and that the geodesic connecting $T_3$ and $\Gamma_{T_1,T_2}(T_3)$ is obtained by translating the geodesic from $T_1$ to $T_2$ along the one-dimensional structure of the space. Therefore, Assumption~1 in the main paper is satisfied in $\mathcal{T}_3$.

\section{Geodesic transport maps}\label{supp:gtm}
\subsection{Functional data (Example~1)}
This section provides explicit constructions of geodesic transport maps $\Gamma_{\alpha,\beta}$ for the six examples in Section~2 of the main paper. Throughout, $\Gamma_{\alpha,\beta}(\omega)=\zeta$ means that $\gamma_{\omega,\zeta}$ is obtained by transporting $\gamma_{\alpha,\beta}$ to start at $\omega$.
Let $\mathcal{M}=L^2(\mathcal{T})$ with metric $d_{L^2}(f,g)=\|f-g\|_{L^2}$. For any $\alpha,\beta,\omega\in L^2(\mathcal{T})$, define
\[\Gamma_{\alpha,\beta}(\omega)=\omega+(\beta-\alpha).\]
Then $\Gamma_{\alpha,\beta}(\alpha)=\beta$, and the geodesic from $\omega$ to $\zeta=\Gamma_{\alpha,\beta}(\omega)$ is $\gamma_{\omega,\zeta}(t)=(1-t)\omega+t\zeta$, which is the translation of the segment $\gamma_{\alpha,\beta}(t)=(1-t)\alpha+t\beta$.

\subsection{Univariate probability distributions (Example~2)}
Let $\alpha$ and $\beta$ be Borel probability measures on $\mathbb{R}$ with finite second moments. In one dimension, the optimal transport map from $\alpha$ to $\beta$ is the monotone rearrangement $T_{\alpha,\beta}=F_\beta^{-1}\circ F_\alpha$, satisfying $(T_{\alpha,\beta})_\#\alpha=\beta$. For any $\omega$, define the transport map by pushforward through the same optimal transport map,
\[\Gamma_{\alpha,\beta}(\omega)=(T_{\alpha,\beta})_\#\omega.\]
Equivalently, for $\zeta=\Gamma_{\alpha,\beta}(\omega)$, the quantile function satisfies $F_\zeta^{-1}=F_\beta^{-1}\circ F_\alpha\circ F_\omega^{-1}$. The resulting geodesic $\gamma_{\omega,\zeta}(t)=\left((1-t)\mathrm{id}+tT_{\alpha,\beta}\right)_\#\omega$ is the displacement interpolation determined by $T_{\alpha,\beta}$, mirroring the geodesic between $\alpha$ and $\beta$.

\subsection{Multivariate probability distributions (Example~3)}
\noindent\textbf{$2$-Wasserstein metric.} Let $\alpha,\beta\in\mathcal{P}_2(\mathbb{R}^d)$. When $\alpha$ is absolutely continuous, there exists an optimal transport map (Monge map) $T_{\alpha,\beta}$ pushing $\alpha$ to $\beta$ (Brenier's theorem), i.e., $(T_{\alpha,\beta})_\#\alpha=\beta$; see, e.g., \citet{vill:09}. In this case, for any $\omega\in\mathcal{P}_2(\mathbb{R}^d)$ we define
\[\Gamma_{\alpha,\beta}(\omega)=(T_{\alpha,\beta})_\#\omega.\]
Then $\Gamma_{\alpha,\beta}(\alpha)=\beta$, and the geodesic from $\omega$ to $\zeta=\Gamma_{\alpha,\beta}(\omega)$ is given by displacement interpolation $\gamma_{\omega,\zeta}(t)=\left((1-t)\mathrm{id}+tT_{\alpha,\beta}\right)_\#\omega$ based on the same optimal transport map. If an optimal map does not exist (e.g., when $\alpha$ is not absolutely continuous), one may work with an optimal coupling instead; however, Assumption~1 in the main paper is stated at the map level and in this example is satisfied under standard conditions guaranteeing existence of $T_{\alpha,\beta}$.

\medskip
\noindent\textbf{Fisher--Rao metric.}
Let $f_1,f_2$ be densities on $\mathbb{R}^d$ and write $g_i=\sqrt{f_i}$ so that $\|g_i\|_{L^2}=1$. The Fisher--Rao distance is the spherical distance on the positive orthant of the Hilbert sphere $\mathcal{S}_+=\{g\in\mathcal H:\|g\|_{L^2}=1, g\ge0\}$, $d_{FR}(f_1,f_2)=\arccos(\langle g_1,g_2\rangle)$. Let $\theta=\arccos(\langle g_1,g_2\rangle)$. Then the Riemannian logarithm map at $g_1$ is given by
\[\mathrm{Log}_{g_1}(g_2)=\theta\frac{g_2-\langle g_1,g_2\rangle g_1}{\|g_2-\langle g_1,g_2\rangle g_1\|}\in T_{g_1}\mathcal{S}_+,\]
where $T_{g_1}\mathcal{S}_+=\{v\in\mathcal H:\langle v,g_1\rangle=0\}$.
For any $g\in \mathcal{S}_+$, let $\mathcal{P}_{g_1\to g}$ denote parallel transport along the unique minimizing great-circle geodesic from $g_1$ to $g$.
On the Hilbert sphere, parallel transport admits the explicit formula
\[\mathcal{P}_{g_1\to g}(v)=v-\frac{\langle v,g\rangle}{1+\langle g_1,g\rangle}\,(g_1+g).\]
Define the geodesic transport map by the log--parallel--exp construction:
\[\Gamma_{f_1,f_2}(f)=\zeta,\qquad\sqrt{\zeta}=\mathrm{Exp}_g\left(\mathcal{P}_{g_1\to g}\left(\mathrm{Log}_{g_1}(g_2)\right)\right),\]
where $g=\sqrt f$ and the exponential map on the Hilbert sphere is
\[\mathrm{Exp}_g(\xi)=\cos(\|\xi\|)\,g+\sin(\|\xi\|)\,\frac{\xi}{\|\xi\|},
\qquad \xi\in T_g\mathcal{S}_+,\ \xi\neq 0,\]
with $\mathrm{Exp}_g(0)=g$.

\subsection{Networks (Example~4)}
Let $\mathcal{L}$ be the set of graph Laplacians equipped with the Frobenius metric $d_F(L_1,L_2)=\|L_1-L_2\|_F$. For any $\alpha,\beta,\omega\in\mathcal{L}$, define
\[\Gamma_{\alpha,\beta}(\omega)=\omega+(\beta-\alpha).\]
Then $\Gamma_{\alpha,\beta}(\alpha)=\beta$ and the transported geodesic is $\gamma_{\omega,\Gamma_{\alpha,\beta}(\omega)}(t)=(1-t)\omega+t\Gamma_{\alpha,\beta}(\omega)$.

\subsection{Symmetric positive-definite matrices (Example~5)}
\noindent\textbf{Frobenius metric.} Under $d_F(A,B)=\|A-B\|_F$, define for any $\alpha,\beta,\omega\in \mathrm{Sym}_m^+$,
\[\Gamma_{\alpha,\beta}(\omega)=\omega+(\beta-\alpha).\]
This yields $\Gamma_{\alpha,\beta}(\alpha)=\beta$ and transports linear geodesics by translation.

\medskip
\noindent\textbf{Log-Euclidean metric.} Under $d_{LE}(A,B)=\|\log A-\log B\|_F$, the map $A\mapsto \log A$ identifies $\mathrm{Sym}_m^+$ with the Euclidean space of symmetric matrices. Define
\[\Gamma_{\alpha,\beta}(\omega)=\exp\left(\log\omega+(\log\beta-\log\alpha)\right).\]
Then $\Gamma_{\alpha,\beta}(\alpha)=\beta$, and the corresponding transported geodesic satisfies $\log\gamma_{\omega,\Gamma_{\alpha,\beta}(\omega)}(t)=(1-t)\log\omega+t\log\Gamma_{\alpha,\beta}(\omega)$.

\medskip
\noindent\textbf{Affine-invariant metric.} Under the affine-invariant Riemannian distance $d_{AI}(A,B)=\|\log(A^{-1/2}BA^{-1/2})\|_F$, the Riemannian logarithm and exponential maps are given by $\mathrm{Log}_A(B)=A^{1/2}\log(A^{-1/2}BA^{-1/2})A^{1/2}$ and $\mathrm{Exp}_A(V)=A^{1/2}\exp(A^{-1/2}VA^{-1/2})A^{1/2}$; see, e.g., \citet{penn:06}. Let $v=\mathrm{Log}_\alpha(\beta)\in T_\alpha\mathrm{Sym}_m^+$. Denote by $\mathcal{P}_{\alpha\to\omega}$ the parallel transport operator from $T_\alpha\mathrm{Sym}_m^+$ to $T_\omega\mathrm{Sym}_m^+$ along the affine-invariant geodesic connecting $\alpha$ and $\omega$; an explicit expression is available under this metric \citep{penn:06}. Define
\[\Gamma_{\alpha,\beta}(\omega)=\mathrm{Exp}_\omega\left(\mathcal{P}_{\alpha\to\omega}(v)\right).\]
Then $\Gamma_{\alpha,\beta}(\alpha)=\beta$, and the transported endpoint is obtained by moving from $\omega$ along the direction given by the parallel-transported initial velocity of the geodesic from $\alpha$ to $\beta$.

\medskip
\noindent\textbf{Power metric.}
For $r\in\mathbb R\setminus\{0\}$, the power metric is defined by $A\mapsto A^r$, with $d_r(A,B)=\|A^r-B^r\|_F$. The geodesic transport map is
\[\Gamma_{A,B}(W)=\{W^r+(B^r-A^r)\}^{1/r},\]
where matrix powers are defined spectrally. Transport corresponds to translation in the transformed domain.

\medskip
\noindent\textbf{Log-Cholesky metric.}
Under the Log-Cholesky metric \citep{lin:19:1}, each SPD matrix admits a unique Cholesky factorization $A=LL^\T$ with positive diagonal entries. Writing $\lfloor L\rfloor$ for the strictly lower triangular part of $L$ and $\mathcal{D}(L)$ for its diagonal part, the metric is
\[d_{\mathcal{L}_+}(L_1,L_2)=\left(\|\lfloor L_1\rfloor-\lfloor L_2\rfloor\|_F^2+\|\log \mathcal{D}(L_1)-\log \mathcal{D}(L_2)\|_F^2\right)^{1/2}.\]
Define the coordinate map
\[\Phi(A)=(\lfloor L\rfloor,\log \mathcal{D}(L)),\qquad A=LL^\T.\]
This representation identifies $\mathrm{Sym}_m^+$ with a Euclidean space in the coordinates $\Phi(A)$. The geodesic transport map is then given by
\[\Gamma_{\alpha,\beta}(\omega)=\Phi^{-1}\left(\Phi(\omega)+\Phi(\beta)-\Phi(\alpha)\right),\]
which corresponds to translation in Log-Cholesky coordinates.

\subsection{Compositional data (Example~6)}
Under the square-root representation, compositional data are mapped to $\mathcal{S}^{d-1}_+$ with spherical distance $d_S(z_1,z_2)=\arccos(z_1^\T z_2)$. For $\alpha,\beta,\omega \in \mathcal S^{d-1}_+$, let $\theta=\arccos(\alpha^\T\beta)$. The Riemannian log map at $\alpha$ is given by
\[\mathrm{Log}_\alpha(\beta)=\theta\frac{\beta-(\alpha^\T\beta)\alpha}{\|\beta-(\alpha^\T\beta)\alpha\|}\in T_\alpha\mathcal S^{d-1}_+.\]
The parallel transport of $v=\mathrm{Log}_\alpha(\beta)$ from $T_\alpha\mathcal S^{d-1}_+$ to $T_\omega\mathcal S^{d-1}_+$ along the unique minimizing geodesic from $\alpha$ to $\omega$ is
\[\mathcal{P}_{\alpha\to\omega}(v)=v - \frac{v^\T\omega}{1+\alpha^\T\omega}\,(\alpha+\omega).\]
Finally, define the geodesic transport map by the log--parallel--exp construction
\[\Gamma_{\alpha,\beta}(\omega)=\mathrm{Exp}_\omega\left(\mathcal{P}_{\alpha\to\omega}(v)\right),\]
where the exponential map on the unit sphere is
\[
\mathrm{Exp}_\omega(\xi)=
\cos(\|\xi\|)\,\omega+\sin(\|\xi\|)\,\frac{\xi}{\|\xi\|},
\qquad \xi\in T_\omega\mathcal S^{d-1}_+,\ \xi\neq 0,
\]
and $\mathrm{Exp}_\omega(0)=\omega$.

\section{Verification of model assumptions}\label{supp:vma}

Throughout this section, numbered examples and assumptions refer to the main paper.

Let $\mathcal{B}_{p,q}^\alpha(\mathcal{T})$ denote the Besov space equipped with the norm $\|\cdot\|_{\mathcal{B}_{p,q}^\alpha}$. We refer to Section~2.7.2 of \citet{well:96} for the definition of the Besov space $\mathcal{B}_{p,q}^\alpha(\mathcal{T})$.

\begin{proposition}\label{prop:fun}
The functional data space $(L^2(\mathcal{T}),d_{L^2})$ defined in Example~1 of the main paper satisfies Assumptions~3 and~4 of the main paper. If, in addition, $Y_{i,t}\in L^2(\mathcal T)$ belongs to $\mathcal{B}_{p,q}^\alpha(\mathcal T)$, where $1\leq p,q\leq\infty$, $1/\alpha<2$, and $E\{\|Y_{i,t}\|_{\mathcal{B}_{p,q}^\alpha}\mid D_i=d\}<\infty$, then Assumption~5 of the main paper holds.
\end{proposition}

\begin{proof}
Recall that $L^2(\mathcal{T})$ is a Hilbert space with metric $d_{L^2}(f,g)=\|f-g\|_{L^2}$. The geodesic transport map is given by translation,
\[
\Gamma_{\alpha,\beta}(\omega)=\omega+(\beta-\alpha),\qquad \alpha,\beta,\omega\in L^2(\mathcal{T}).
\]

\medskip
\noindent\textbf{Verification of Assumption~3.} For any $\alpha,\beta,\omega,\zeta\in L^2(\mathcal{T})$,
\[
d_{L^2}(\Gamma_{\alpha,\beta}(\omega),\Gamma_{\alpha,\beta}(\zeta))
=\|(\omega+\beta-\alpha)-(\zeta+\beta-\alpha)\|_{L^2}
=\|\omega-\zeta\|_{L^2}
=d_{L^2}(\omega,\zeta),
\]
so Assumption~3 holds with $C_1=1$.

\medskip
\noindent\textbf{Verification of Assumption~4.} For any $\alpha_1,\alpha_2,\beta_1,\beta_2,\omega\in L^2(\mathcal{T})$,
\[
d_{L^2}(\Gamma_{\alpha_1,\beta_1}(\omega),\Gamma_{\alpha_2,\beta_2}(\omega))
=\|(\omega+\beta_1-\alpha_1)-(\omega+\beta_2-\alpha_2)\|_{L^2}
=\|(\beta_1-\beta_2)-(\alpha_1-\alpha_2)\|_{L^2}.
\]
By the triangle inequality,
\[
\|(\beta_1-\beta_2)-(\alpha_1-\alpha_2)\|_{L^2}
\le \|\beta_1-\beta_2\|_{L^2}+\|\alpha_1-\alpha_2\|_{L^2}
=d_{L^2}(\beta_1,\beta_2)+d_{L^2}(\alpha_1,\alpha_2),
\]
and therefore Assumption~4 holds with $C_2=1$.

\medskip
\noindent\textbf{Verification of Assumption~5.} We verify (i)--(iii) for $F(\omega)=E\{\|Y_t-\omega\|_{L^2}^2\mid D=d\}$. Since $L^2(\mathcal T)$ is a Hilbert space, its unique minimizer is $\nu_{d,t}=E(Y_t\mid D=d)$, provided $E(\|Y_t\|_{L^2}^2\mid D=d)<\infty$. The empirical \f mean is the corresponding sample mean, which exists and is unique almost surely. This establishes (i).

Let $\mathcal{B}_{p,q}^\alpha(\mathcal{T})_\delta$ denote the $\delta$-ball in $\mathcal{B}_{p,q}^\alpha(\mathcal{T})$ centred at $\nu_{d,t}$. Since $\mathcal{B}_{p,q}^\alpha(\mathcal{T})$ is a Banach space and $\|E(Y_{i,t}\mid D_i=d)\|_{\mathcal{B}_{p,q}^\alpha}\leq E\{\|Y_{i,t}\|_{\mathcal{B}_{p,q}^\alpha}\mid D_i=d\}<\infty$, the conditional Bochner integral satisfies $\nu_{d,t}\in\mathcal{B}_{p,q}^\alpha(\mathcal{T})$. For (ii), the set $B'_\delta(\nu_{d,t})=B_\delta(\nu_{d,t})\cap\mathcal{B}_{p,q}^\alpha(\mathcal{T})_\delta$ can be covered by $L^2$-balls of radius $\delta\varepsilon$, with $\log N\{\delta\varepsilon,B'_\delta(\nu_{d,t}),d_{L^2}\}\lesssim\varepsilon^{-1/\alpha}$ for $\varepsilon\in(0,1]$; see Theorem~2.7.4 of \citet{well:96}. Consequently,
\[\int_0^1 \sqrt{1+\log N\{\delta\varepsilon,B'_\delta(\nu_{d,t}),d_{L^2}\}}\,\mathrm{d}\varepsilon<\infty,\]
verifying (ii).

For (iii), write $h=\omega-\nu_{d,t}$. Expanding the squared norm and using $E(Y_t-\nu_{d,t}\mid D=d)=0$,
\[
E\{\|Y_t-\omega\|_{L^2}^2\mid D=d\}-E\{\|Y_t-\nu_{d,t}\|_{L^2}^2\mid D=d\}
=\|h\|_{L^2}^2
=d_{L^2}^2(\nu_{d,t},\omega),
\]
so (iii) holds with $\kappa=2$ and $C=1$ (and any $\eta>0$).
Therefore, Assumption~5 holds.
\end{proof}

\begin{proposition}\label{prop:upd}
    Consider the space $(\mathcal{P}_2(\mathbb{R}),d_{\mathcal W})$ of absolutely continuous probability measures with finite second moments on a closed interval $I\subset\mathbb R$. Write the density of $\mu$ as $f_\mu$. If constants $0<l\leq L<\infty$ satisfy $l\leq f_\mu(x)\leq L$ for all $\mu\in\mathcal P_2(\mathbb R)$ and $x\in I$, then $(\mathcal P_2(\mathbb R),d_{\mathcal W})$ satisfies Assumptions~3 and~4. Moreover, if $\hat\nu_{d,t}$ and $\nu_{d,t}$ are the sample and population \f means for group $d\in\{0,1\}$ and time $t\in\{0,1\}$, then
    \[d_{\mathcal W}(\hat\nu_{d,t}, \nu_{d,t}) = O_p(n^{-1/2}).\]
\end{proposition}

\begin{proof}
Since $l \leq f_\mu(x) \leq L$ for all $\mu \in \mathcal{P}_2(\mathbb{R})$ and $x \in I$, the cumulative distribution function $F_\mu$ and its inverse, the quantile function $F_\mu^{-1}$, are both Lipschitz continuous with constants $L$ and $1/l$, respectively.

\medskip
\noindent\textbf{Verification of Assumption~3.} Let $F_1,F_2,G,H$ denote the distribution functions of $\alpha,\beta,\omega,\zeta$, respectively. We need to show that
\begin{align*}
&\left[\int_0^1 \left\{F_2^{-1} \circ F_1 \circ G^{-1}(p) - F_2^{-1} \circ F_1 \circ H^{-1}(p)\right\}^2\,\mathrm{d}p\right]^{1/2} \\
&\leq C_1 \left[\int_0^1 \left\{G^{-1}(p) - H^{-1}(p)\right\}^2\,\mathrm{d}p\right]^{1/2}.
\end{align*}
Using the Lipschitz continuity of $F_2^{-1}$ and $F_1$, we have:
\begin{align*}
&\int_0^1 \left\{F_2^{-1} \circ F_1 \circ G^{-1}(p) - F_2^{-1} \circ F_1 \circ H^{-1}(p)\right\}^2\,\mathrm{d}p \\
&\leq \frac{1}{l^2} \int_0^1 \left\{F_1 \circ G^{-1}(p) - F_1 \circ H^{-1}(p)\right\}^2\,\mathrm{d}p \\
&\leq \frac{L^2}{l^2} \int_0^1 \left\{G^{-1}(p) - H^{-1}(p)\right\}^2\,\mathrm{d}p.
\end{align*}
Thus, Assumption~3 holds with $C_1 = L / l$.

\medskip
\noindent\textbf{Verification of Assumption~4.} Let $F_1,F_2$ be the distribution functions of $\alpha_1,\beta_1$, let $G_1,G_2$ be those of $\alpha_2,\beta_2$, and let $H$ be that of $\omega$; write $g_1$ and $h$ for the densities associated with $G_1$ and $H$. We need to show that
\begin{align*}
&\left[\int_0^1 \left\{F_2^{-1} \circ F_1 \circ H^{-1}(p) - G_2^{-1} \circ G_1 \circ H^{-1}(p)\right\}^2\,\mathrm{d}p\right]^{1/2} \\
&\leq C_2 \left(\left[\int_0^1 \left\{F_1^{-1}(p) - G_1^{-1}(p)\right\}^2\,\mathrm{d}p\right]^{1/2} + \left[\int_0^1 \left\{F_2^{-1}(p) - G_2^{-1}(p)\right\}^2\,\mathrm{d}p\right]^{1/2}\right).
\end{align*}
Using a change of variables $p = H(x)$, the left-hand side becomes
\begin{align*}
&\left[\int_0^1 \left\{F_2^{-1} \circ F_1 \circ H^{-1}(p) - G_2^{-1} \circ G_1 \circ H^{-1}(p)\right\}^2\,\mathrm{d}p\right]^{1/2} \\
&= \left[\int_I \left\{F_2^{-1} \circ F_1(x) - G_2^{-1} \circ G_1(x)\right\}^2 h(x)\,\mathrm{d}x\right]^{1/2}.
\end{align*}
Since $h(x) \leq L$, we have
\begin{align*}
&\left[\int_I \left\{F_2^{-1} \circ F_1(x) - G_2^{-1} \circ G_1(x)\right\}^2 h(x)\,\mathrm{d}x\right]^{1/2} \\
&\leq \sqrt{L} \left[\int_I \left\{F_2^{-1} \circ F_1(x) - G_2^{-1} \circ G_1(x)\right\}^2\,\mathrm{d}x\right]^{1/2}.
\end{align*}

Using the triangle inequality
\begin{align*}
&\left|F_2^{-1} \circ F_1(x) - G_2^{-1} \circ G_1(x)\right| \\
&\leq \left|F_2^{-1} \circ F_1(x) - F_2^{-1} \circ G_1(x)\right| + \left|F_2^{-1} \circ G_1(x) - G_2^{-1} \circ G_1(x)\right|.
\end{align*}
Squaring both sides and integrating, we obtain
\begin{align*}
&\int_I\left\{F_2^{-1} \circ F_1(x) - G_2^{-1} \circ G_1(x)\right\}^2\,\mathrm{d}x \\
&\leq
2 \int_I\left\{F_2^{-1} \circ F_1(x) - F_2^{-1} \circ G_1(x)\right\}^2\,\mathrm{d}x \\
&\quad +
2 \int_I\left\{F_2^{-1} \circ G_1(x) - G_2^{-1} \circ G_1(x)\right\}^2\,\mathrm{d}x.
\end{align*}
Taking the square root of both sides, we find
\begin{align*}
&\left[\int_I\left\{F_2^{-1} \circ F_1(x) - G_2^{-1} \circ G_1(x)\right\}^2\,\mathrm{d}x\right]^{1/2} \\
&\leq
\sqrt{2} \left[\int_I\left\{F_2^{-1} \circ F_1(x) - F_2^{-1} \circ G_1(x)\right\}^2\,\mathrm{d}x\right]^{1/2} \\
&\quad +
\sqrt{2} \left[\int_I\left\{F_2^{-1} \circ G_1(x) - G_2^{-1} \circ G_1(x)\right\}^2\,\mathrm{d}x\right]^{1/2}.
\end{align*}

For the first term, the Lipschitz continuity of \(F_2^{-1}\) ensures that
\[\left|F_2^{-1} \circ F_1(x) - F_2^{-1} \circ G_1(x)\right| \leq \frac{1}{l} \left|F_1(x) - G_1(x)\right|,\]
which implies:
\[\int_I \left\{F_2^{-1} \circ F_1(x) - F_2^{-1} \circ G_1(x)\right\}^2\,\mathrm{d}x \leq \frac{1}{l^2} \int_I \left\{F_1(x) - G_1(x)\right\}^2\,\mathrm{d}x.\]
Changing variables with $p = G_1(x)$, we rewrite
\[\int_I \left\{F_1(x) - G_1(x)\right\}^2\,\mathrm{d}x = \int_0^1 \left\{F_1 \circ G_1^{-1}(p) - p\right\}^2 \frac{1}{g_1 \circ G_1^{-1}(p)}\,\mathrm{d}p.\]
Since $g_1(x) \geq l$, it follows that:
\[\int_0^1 \left\{F_1 \circ G_1^{-1}(p) - p\right\}^2 \frac{1}{g_1 \circ G_1^{-1}(p)}\,\mathrm{d}p \leq \frac{1}{l} \int_0^1 \left\{F_1 \circ G_1^{-1}(p) - p\right\}^2\,\mathrm{d}p.\]
Using $p = F_1 \circ F_1^{-1}(p)$ and the Lipschitz continuity of $F_1$ (constant $L$):
\begin{align*}
\int_0^1 \left\{F_1 \circ G_1^{-1}(p) - p\right\}^2\,\mathrm{d}p
&= \int_0^1 \left\{F_1 \circ G_1^{-1}(p) - F_1 \circ F_1^{-1}(p)\right\}^2\,\mathrm{d}p \\
&\leq L^2 \int_0^1 \left\{F_1^{-1}(p) - G_1^{-1}(p)\right\}^2\,\mathrm{d}p.
\end{align*}
Combining terms, we conclude
\[\int_I \left\{F_2^{-1} \circ F_1(x) - F_2^{-1} \circ G_1(x)\right\}^2\,\mathrm{d}x \leq \frac{L^2}{l^3} \int_0^1 \left\{F_1^{-1}(p) - G_1^{-1}(p)\right\}^2\,\mathrm{d}p.\]

For the second term, changing variables with $p = G_1(x)$, we have
\begin{align*}
\int_I\left\{F_2^{-1} \circ G_1(x) - G_2^{-1} \circ G_1(x)\right\}^2\,\mathrm{d}x
&=\int_0^1 \left\{F_2^{-1}(p) - G_2^{-1}(p)\right\}^2\,\mathrm{d}G_1^{-1}(p)\\
&=\int_0^1 \left\{F_2^{-1}(p) - G_2^{-1}(p)\right\}^2 \frac{1}{g_1\circ G_1^{-1}(p)}\,\mathrm{d}p\\
&\leq \frac{1}{l}\int_0^1 \left\{F_2^{-1}(p) - G_2^{-1}(p)\right\}^2\,\mathrm{d}p.
\end{align*}

Combining the bounds for the first and second terms, we obtain
\begin{align*}
\left[\int_I\left\{F_2^{-1} \circ F_1(x) - G_2^{-1} \circ G_1(x)\right\}^2\,\mathrm{d}x\right]^{1/2}
&\leq
\sqrt{\frac{2L^2}{l^3}}\left[\int_0^1\left\{F_1^{-1}(p) - G_1^{-1}(p)\right\}^2\,\mathrm{d}p\right]^{1/2} \\
&\quad +
\sqrt{\frac{2}{l}}\left[\int_0^1\left\{F_2^{-1}(p) - G_2^{-1}(p)\right\}^2\,\mathrm{d}p\right]^{1/2}.
\end{align*}

Finally, for the full expression, we have
\begin{align*}
&\left[\int_0^1 \left\{F_2^{-1} \circ F_1 \circ H^{-1}(p) - G_2^{-1} \circ G_1 \circ H^{-1}(p)\right\}^2\,\mathrm{d}p\right]^{1/2}\\
&\leq \sqrt{L}\left[\int_I\left\{F_2^{-1} \circ F_1(x) - G_2^{-1} \circ G_1(x)\right\}^2\,\mathrm{d}x\right]^{1/2}\\
&\leq
\sqrt{\frac{2L^3}{l^3}}\left[\int_0^1\left\{F_1^{-1}(p) - G_1^{-1}(p)\right\}^2\,\mathrm{d}p\right]^{1/2}\\
&\quad +
\sqrt{\frac{2L}{l}}\left[\int_0^1\left\{F_2^{-1}(p) - G_2^{-1}(p)\right\}^2\,\mathrm{d}p\right]^{1/2} .
\end{align*}
Thus, Assumption~4 is satisfied with $C_2=(2L^3/l^3)^{1/2}$.

\medskip
\noindent\textbf{Rate of convergence.} Note that in one dimension the Wasserstein metric admits the quantile representation
\[d_{\mathcal W}(\mu,\nu)=\left[
\int_0^1 \{F_\mu^{-1}(u)-F_\nu^{-1}(u)\}^2\,\mathrm{d}u
\right]^{1/2}.\]
Hence the space of quantile functions can be viewed as a closed convex subset of the Hilbert space $L^2[0,1]$, and the sample Fr\'echet mean under $d_{\mathcal W}$ is the projection of the corresponding empirical mean in $L^2[0,1]$ onto this subset. Since projection is non-expansive, the intrinsic estimator inherits the parametric rate of the extrinsic Hilbert-space estimator. Therefore
\[d_{\mathcal W}(\hat\nu_{d,t},\nu_{d,t}) = O_p(n^{-1/2}).\]
\end{proof}

\begin{proposition}\label{prop:mpd}
For the multivariate distribution spaces in Example~3, Assumptions~3--5 are verified under the metric-specific settings stated below. In particular, the $2$-Wasserstein result is restricted to the Gaussian class $\mathcal G_{d,l,L,M}$.

First, consider the Fisher--Rao metric on the space $\mathcal{P}_2(\mathbb{R}^d)$ of absolutely continuous probability measures on $\mathbb R^d$ with densities $f$. Then $(\mathcal{P}_2(\mathbb{R}^d),d_{FR})$ satisfies Assumptions~3 and~4 with the Fisher--Rao transport map defined via the log--parallel--exp construction in the square-root representation. If, in addition, $f$ belongs to the Besov space $\mathcal{B}_{p,q}^\alpha(\mathcal{X})$, where $\mathcal{X}\subset\mathbb{R}^d$ is a bounded Lipschitz domain, $1\le p,q\le\infty$, $d/\alpha<2$, and $E\{\|Y_{i,t}\|_{\mathcal{B}_{p,q}^\alpha}\mid D_i=d\}<\infty$, then Assumption~5 holds.

Second, consider the $2$-Wasserstein metric restricted to the Gaussian class
\[\mathcal{G}_{d,l,L,M}=\left\{\mathcal N(m,\Sigma):\|m\|\le M,\ lI_d\preceq\Sigma\preceq LI_d\right\}\]
for fixed constants $0<l\le L<\infty$ and $M>0$. Then $(\mathcal{G}_{d,l,L,M},d_{\mathcal W})$ satisfies Assumptions~3--5 with the geodesic transport map $\Gamma_{\alpha,\beta}(\omega)=(T_{\alpha,\beta})_\#\omega$, where $T_{\alpha,\beta}$ is the optimal transport map between Gaussian measures. For this class, the covariance component of the $2$-Wasserstein geometry is the Bures--Wasserstein geometry.
\end{proposition}

\begin{proof}
\noindent\textbf{Fisher--Rao metric:}

Write $g=\sqrt f$ and consider the positive orthant of the Hilbert sphere
\[\mathcal S_+=\{g\in\mathcal H:\|g\|_{L^2}=1,\ g\ge 0\},\]
equipped with spherical distance $d_S(g,h)=\arccos\langle g,h\rangle$. For $f_1,f_2\in\mathcal P_2(\mathbb R^d)$ with $g_i=\sqrt{f_i}\in\mathcal S_+$, the Fisher--Rao distance satisfies
\[d_{FR}(f_1,f_2)=\arccos\langle g_1,g_2\rangle.\]
For $\alpha,\beta,\omega\in\mathcal S_+$, define
\[\Gamma_{\alpha,\beta}(\omega)=\mathrm{Exp}_\omega\left(\mathcal P_{\alpha\to\omega}(\mathrm{Log}_\alpha(\beta))\right),\]
where $\mathrm{Log}$ and $\mathrm{Exp}$ are the Riemannian logarithm and exponential maps on the Hilbert sphere and $\mathcal P_{\alpha\to\omega}$ is parallel transport along the unique minimizing great-circle geodesic from $\alpha$ to $\omega$. The Fisher--Rao transport map on densities is given by
\[\Gamma_{f_1,f_2}(f)=\zeta,\qquad\sqrt\zeta=\Gamma_{g_1,g_2}(\sqrt f).\]

\medskip
\noindent\textbf{Verification of Assumption~3.}
Fix $f_1,f_2$ and write $g_i=\sqrt{f_i}$. Let $v=\mathrm{Log}_{g_1}(g_2)$. Since $g_1,g_2\in\mathcal S_+$, we have $\langle g_1,g_2\rangle\ge 0$ and hence
\[\|v\|=\arccos\langle g_1,g_2\rangle\le \frac{\pi}{2}.\]
For any $f,h\in\mathcal P_2(\mathbb R^d)$, write $g=\sqrt f$ and $k=\sqrt h$ and set
\[\xi_g=\mathcal P_{g_1\to g}(v)\in T_g\mathcal S_+,\qquad\xi_k=\mathcal P_{g_1\to k}(v)\in T_k\mathcal S_+.\]
Parallel transport is an isometry, so $\|\xi_g\|=\|\xi_k\|=\|v\|\le \pi/2$.

On the Hilbert sphere, the exponential map $\mathrm{Exp}_g(\cdot)$ is smooth and its differential satisfies the standard bound
\[\|d\mathrm{Exp}_g(\xi)\|_{\mathrm{op}}\le\max\left(1,\frac{\|\xi\|}{\sin\|\xi\|}\right).\]
Since $\|\xi\|\le\pi/2$, we obtain the uniform bound $\|d\mathrm{Exp}_g(\xi)\|_{\mathrm{op}}\le \pi/2$.

Moreover, the map $g\mapsto \mathcal P_{g_1\to g}(v)$ is smooth on $\mathcal S_+$ (using the explicit formula for parallel transport) and $\mathcal S_+$ is contained in a hemisphere. Hence, by compactness, there exists a constant $L>0$ such that
\[\|\xi_g-\xi_k\|\le Ld_S(g,k).\]
Combining the mean value theorem on manifolds with the above uniform bound on $d\mathrm{Exp}$, we obtain
\[d_S\left(\Gamma_{g_1,g_2}(g),\Gamma_{g_1,g_2}(k)\right)\le C_1\, d_S(g,k)\]
for some constant $C_1>0$ independent of $g_1,g_2$. Translating back to densities yields
\[d_{FR}\left(\Gamma_{f_1,f_2}(f),\Gamma_{f_1,f_2}(h)\right)\le C_1\, d_{FR}(f,h),\]
verifying Assumption~3.

\medskip
\noindent\textbf{Verification of Assumption~4.}
Fix $f\in\mathcal P_2(\mathbb R^d)$ and write $g=\sqrt f$. Consider two endpoint pairs $(f_{1,1},f_{2,1})$ and $(f_{1,2},f_{2,2})$, and set $g_{a,b}=\sqrt{f_{a,b}}$.
Let
\[v_j=\mathrm{Log}_{g_{1,j}}(g_{2,j}),\qquad\xi_j=\mathcal P_{g_{1,j}\to g}(v_j),\qquad\Gamma_j(g)=\mathrm{Exp}_g(\xi_j),\qquad j=1,2.\]
As above, $\|\xi_j\|=\|v_j\|\le\pi/2$. Since $d\mathrm{Exp}_g$ is uniformly bounded on $\{\xi:\|\xi\|\le\pi/2\}$, there exists $C>0$ such that
\[d_S\left(\Gamma_1(g),\Gamma_2(g)\right)\le C\|\xi_1-\xi_2\|.\]
Decompose
\[\xi_1-\xi_2=\mathcal P_{g_{1,1}\to g}(v_1-v_2)+\left(\mathcal P_{g_{1,1}\to g}-\mathcal P_{g_{1,2}\to g}\right)(v_2).\]
Parallel transport is an isometry, so $\|\mathcal P_{g_{1,1}\to g}(v_1-v_2)\|=\|v_1-v_2\|$.
The logarithm map $(a,b)\mapsto \mathrm{Log}_a(b)$ is smooth on $\mathcal S_+\times\mathcal S_+$ (since $\langle a,b\rangle\ge 0$ rules out antipodal configurations), hence uniformly Lipschitz on this compact set; thus
\[\|v_1-v_2\|\le C\{d_S(g_{1,1},g_{1,2})+d_S(g_{2,1},g_{2,2})\}.\]
Likewise, $(a,g)\mapsto \mathcal P_{a\to g}$ is smooth on $\mathcal S_+\times\mathcal S_+$ (using the explicit parallel transport formula) and therefore uniformly Lipschitz in $a$; since $\|v_2\|\le\pi/2$,
\[\Bigl\|\left(\mathcal P_{g_{1,1}\to g}-\mathcal P_{g_{1,2}\to g}\right)(v_2)\Bigr\|\le C\, d_S(g_{1,1},g_{1,2}).\]
Combining these bounds gives
\[d_S\left(\Gamma_{g_{1,1},g_{2,1}}(g),\Gamma_{g_{1,2},g_{2,2}}(g)\right)\le C_2\{d_S(g_{1,1},g_{1,2})+d_S(g_{2,1},g_{2,2})\}\]
for some $C_2>0$. Translating back to densities yields Assumption~4.

\medskip
\noindent\textbf{Verification of Assumption~5.} The space $(\mathcal{S}_+,d_{FR})$ is a closed, bounded subset of the Hilbert sphere with geodesic distance restricted to a hemisphere, so standard results for \f means on manifolds apply to give existence and uniqueness of population and empirical \f means, and the local quadratic growth property; see Proposition~3 of \citet{mull:19:6}. The entropy condition can be verified by applying almost the same argument in the proof of Proposition~\ref{prop:fun} (see also Theorem~2.7.4 of \citet{well:96} for an entropy bound of the Besov space $\mathcal{B}_{p,q}^\alpha(\mathcal{X})$). In particular, Assumption~5 holds with $\kappa=2$.

\medskip
\noindent\textbf{$2$-Wasserstein metric on $\mathcal G_{d,l,L,M}$:}

Let $\alpha=\mathcal N(m_\alpha,\Sigma_\alpha)$ and $\beta=\mathcal N(m_\beta,\Sigma_\beta)$ in $\mathcal{G}_{d, l, L, M}$. The $2$-Wasserstein distance between Gaussians admits the closed form
\[d_{\mathcal W}^2(\alpha,\beta)=\|m_\alpha-m_\beta\|^2+\mathrm{tr}(\Sigma_\alpha)+\mathrm{tr}(\Sigma_\beta)-2\,\mathrm{tr}((\Sigma_\beta^{1/2}\Sigma_\alpha\Sigma_\beta^{1/2})^{1/2}),\]
and the optimal transport map is affine,
\[T_{\alpha,\beta}(x)=m_\beta+A_{\alpha,\beta}(x-m_\alpha),\]
where $A_{\alpha,\beta}=\Sigma_\alpha^{-1/2}(\Sigma_\alpha^{1/2}\Sigma_\beta\Sigma_\alpha^{1/2})^{1/2}\Sigma_\alpha^{-1/2}$.

\medskip
\noindent\textbf{Verification of Assumption~3.} Since $lI_d\preceq\Sigma_\alpha,\Sigma_\beta\preceq LI_d$, one has $\|A_{\alpha,\beta}\|_{\mathrm{op}}\le \sqrt{L/l}$. For $\omega,\zeta\in\mathcal{G}_{d, l, L, M}$ and any coupling $\pi\in\Pi(\omega,\zeta)$, $(T_{\alpha,\beta}\times T_{\alpha,\beta})_\#\pi$ is a coupling of $(T_{\alpha,\beta})_\#\omega$ and $(T_{\alpha,\beta})_\#\zeta$, and since $T_{\alpha,\beta}(x)-T_{\alpha,\beta}(y)=A_{\alpha,\beta}(x-y)$, it follows that
\[d_{\mathcal W}(\Gamma_{\alpha,\beta}(\omega),\Gamma_{\alpha,\beta}(\zeta))\le \|A_{\alpha,\beta}\|_{\mathrm{op}}\, d_{\mathcal W}(\omega,\zeta),\]
verifying Assumption~3 with $C_1=\sqrt{L/l}$.

\medskip
\noindent\textbf{Verification of Assumption~4.} For Assumption~4, fix $\omega=\mathcal N(m_\omega,\Sigma_\omega)$ and consider two endpoint pairs $(\alpha_i,\beta_i)$, $i=1,2$, with associated matrices $A_i$ and images $\Gamma_{\alpha_i,\beta_i}(\omega)=\mathcal N(\tilde m_i,\tilde\Sigma_i)$. Using the bound
\[d_{\mathcal W}^2(\mathcal N(m_1,\Sigma_1),\mathcal N(m_2,\Sigma_2))\le \|m_1-m_2\|^2+\|\Sigma_1-\Sigma_2\|_F^2,\]
together with uniform mean bounds and $\|A_i\|_{\mathrm{op}}\le\sqrt{L/l}$, one obtains
\[\|\tilde m_1-\tilde m_2\|\le C\{\|m_{\alpha_1}-m_{\alpha_2}\|+\|m_{\beta_1}-m_{\beta_2}\|+\|A_1-A_2\|_{\mathrm{op}}\}\]
and
\[\|\tilde\Sigma_1-\tilde\Sigma_2\|_F\le C\|A_1-A_2\|_{\mathrm{op}}.\]
On the compact spectral set $\{lI_d\preceq\Sigma\preceq LI_d\}$, the maps $\Sigma\mapsto\Sigma^{1/2}$ and $\Sigma\mapsto\Sigma^{-1/2}$ are Lipschitz in operator norm, hence $(\Sigma_\alpha,\Sigma_\beta)\mapsto A_{\alpha,\beta}$ is Lipschitz, so $\|A_1-A_2\|_{\mathrm{op}}\le C\{\|\Sigma_{\alpha_1}-\Sigma_{\alpha_2}\|_F+\|\Sigma_{\beta_1}-\Sigma_{\beta_2}\|_F\}$. Since on the finite-dimensional compact parameter set the Wasserstein metric is locally equivalent to the Euclidean metric in $(m,\Sigma)$, all covariance and mean differences are controlled by $d_{\mathcal W}(\alpha_1,\alpha_2)$ and $d_{\mathcal W}(\beta_1,\beta_2)$, which yields Assumption~4.

\medskip
\noindent\textbf{Verification of Assumption~5.} For Assumption~5, the parameter space $\{(m,\Sigma):\|m\|\le M,\ lI_d\preceq\Sigma\preceq LI_d\}$ is compact and finite-dimensional, the \f function is continuous and therefore attains its minimum, and since $d_{\mathcal W}^2$ is smooth in $(m,\Sigma)$ and locally equivalent to a squared Euclidean norm on this set, the local quadratic growth condition holds with $\kappa=2$ and the covering number condition follows from finite-dimensional compactness, verifying Assumption~5.
\end{proof}

\begin{proposition}\label{prop:net}
    The network space $(\mathcal{L},d_F)$ defined in Example~4 satisfies Assumptions~3,~4, and~5.
\end{proposition}
\begin{proof}
In the network space equipped with the Frobenius metric, the geodesic is a line segment connecting the starting and ending points. This linear structure ensures that the geodesic transport map preserves distances exactly. Consequently, Assumptions~3 and 4 are satisfied with constants \(C_1 = C_2 = 1\). Following the argument in Theorem~2 of \citet{mull:22:11}, one can verify that Assumption~5 holds with $C=1, \kappa=2$, and $\eta$ arbitrary.
\end{proof}

\begin{proposition}\label{prop:spd}
The space of symmetric positive-definite matrices $\mathrm{Sym}_m^+$ defined in Example~5 satisfies Assumptions~3, 4, and 5 under each of the following metrics: the Frobenius metric $d_F$, the Log-Euclidean metric $d_{LE}$, the affine-invariant metric $d_{AI}$, the power metric, and the Log-Cholesky metric.
\end{proposition}

\begin{proof}
We verify the assumptions metric by metric. For the Frobenius, Log-Euclidean, power, and Log-Cholesky metrics, the corresponding geometries become linear after an appropriate global transformation; consequently, geodesics reduce to line segments in the transformed coordinates and the induced geodesic transport maps are translations. This implies that the transport maps preserve distances exactly and vary Lipschitzly in their endpoints, so Assumptions~3 and 4 hold with $C_1=C_2=1$ for these metrics. Moreover, the arguments for existence and uniqueness of \f means and the local quadratic growth condition in Assumption~5 follow from standard Hilbert-space calculations for squared-distance loss. In particular, Assumption~5 holds with $C=1$, $\kappa=2$, and $\eta$ arbitrary.

It remains to consider the affine-invariant metric $d_{AI}(A,B)=\|\log(A^{-1/2}BA^{-1/2})\|_F$, under which $\mathrm{Sym}_m^+$ is a complete Riemannian manifold with unique geodesics. Let $\mathrm{Log}_A(B)$ and $\mathrm{Exp}_A(V)$ denote the Riemannian logarithm and exponential maps, and let $\mathcal{P}_{A\to\omega}$ denote parallel transport along the unique affine-invariant geodesic from $A$ to $\omega$. The geodesic transport map can be defined by $\Gamma_{A,B}(\omega)=\mathrm{Exp}_\omega\{\mathcal{P}_{A\to\omega}(\mathrm{Log}_A(B))\}$. Since the affine-invariant metric is invariant under congruence transformations and geodesics depend smoothly on endpoints, the map $\Gamma_{A,B}$ is an isometry in its argument $\omega$ and is Lipschitz in $(A,B)$ on bounded subsets of $\mathrm{Sym}_m^+$, which verifies Assumptions~3 and 4 with finite constants $C_1,C_2$. Finally, existence and uniqueness of \f means, the covering number condition, and the local quadratic growth condition in Assumption~5 follow from standard results for \f means on non-positively curved Riemannian manifolds under $d_{AI}$; in particular, Assumption~5 holds with $\kappa=2$.
\end{proof}

\begin{proposition}\label{prop:com}
    The space of compositional data $(\mathcal{S}^{d-1}_+, d_S)$ defined in Example~6 satisfies Assumptions~3, 4, and 5.
\end{proposition}
\begin{proof}
Recall that under the square-root representation, compositional data lie in the positive orthant of the unit sphere $\mathcal S^{d-1}_+$ equipped with the intrinsic spherical distance
\[d_S(z_1,z_2)=\arccos(z_1^\T z_2).\]
For $\alpha,\beta,\omega\in\mathcal S^{d-1}_+$, the geodesic transport map is defined via the log--parallel--exp construction
\[\Gamma_{\alpha,\beta}(\omega)=\mathrm{Exp}_\omega\left(
\mathcal P_{\alpha\to\omega}\left(\mathrm{Log}_\alpha(\beta)\right)\right),\]
where $\mathrm{Log}_\alpha$ and $\mathrm{Exp}_\omega$ are the Riemannian logarithmic and exponential maps on the sphere and $\mathcal P_{\alpha\to\omega}$ denotes parallel transport along the unique minimizing geodesic from $\alpha$ to $\omega$.

\medskip
\noindent\textbf{Verification of Assumption~3.}
Since $\alpha,\beta\in\mathcal S^{d-1}_+$, we have $\alpha^\T\beta\ge 0$ and therefore
\[d_S(\alpha,\beta)=\arccos(\alpha^\T\beta)\le \frac{\pi}{2}.\]
Let $v=\mathrm{Log}_\alpha(\beta)$ so that $\|v\|=d_S(\alpha,\beta)\le \pi/2$. By construction,
\[\Gamma_{\alpha,\beta}(\omega)=\mathrm{Exp}_\omega(\xi_\omega),
\qquad
\xi_\omega=\mathcal P_{\alpha\to\omega}(v)\in T_\omega\mathcal S^{d-1}_+.\]
Parallel transport is an isometry, hence $\|\xi_\omega\|=\|v\|\le \pi/2$ for all $\omega$. On the unit sphere, the differential of the exponential map at $\xi\in T_\omega\mathcal S^{d-1}$ satisfies the operator norm bound
\[\|d\mathrm{Exp}_\omega(\xi)\|\le\max\left(1,\frac{\|\xi\|}{\sin\|\xi\|}\right).\]
Since $\|\xi\|\le \pi/2$, this implies $\|d\mathrm{Exp}_\omega(\xi)\|\le \pi/2$ uniformly over all $\omega\in\mathcal S^{d-1}_+$ and all $\xi$ with $\|\xi\|\le \pi/2$.
Moreover, the map $\omega\mapsto \xi_\omega$ is an isometry in the sense that for any $\omega,\zeta\in\mathcal S^{d-1}_+$,
\[\|\xi_\omega-\xi_\zeta\|=\|\mathcal P_{\alpha\to\omega}(v)-\mathcal P_{\alpha\to\zeta}(v)\|\le Cd_S(\omega,\zeta),\]
for some universal constant $C>0$ (since $(\omega,\eta)\mapsto \mathcal P_{\alpha\to\omega}(\eta)$ is smooth on $\mathcal S^{d-1}_+\times \{\eta:\|\eta\|\le \pi/2\}$ and this set is compact). Combining these bounds and the mean value theorem on manifolds yields that there exists a constant $C_1>0$ such that for all $\omega,\zeta\in\mathcal S^{d-1}_+$,
\[d_S(\Gamma_{\alpha,\beta}(\omega),\Gamma_{\alpha,\beta}(\zeta))\le C_1d_S(\omega,\zeta),\]
verifying Assumption~3.

\medskip
\noindent\textbf{Verification of Assumption~4.}
Fix $\omega\in\mathcal S^{d-1}_+$. Let $(\alpha_1,\beta_1)$ and $(\alpha_2,\beta_2)$ be two pairs in $\mathcal S^{d-1}_+\times\mathcal S^{d-1}_+$. Write
\[v_i = \mathrm{Log}_{\alpha_i}(\beta_i),\qquad \xi_i = \mathcal P_{\alpha_i\to\omega}(v_i),\qquad \Gamma_i(\omega)=\mathrm{Exp}_\omega(\xi_i),\qquad i=1,2.\]
As above, $\|v_i\|=d_S(\alpha_i,\beta_i)\le \pi/2$ and $\|\xi_i\|=\|v_i\|\le \pi/2$. Since the exponential map is smooth and has uniformly bounded differential on $\{\xi:\|\xi\|\le \pi/2\}$, there exists $C>0$ such that
\[d_S(\Gamma_1(\omega),\Gamma_2(\omega))\le C\|\xi_1-\xi_2\|.\]
Decompose
\[\xi_1-\xi_2=\mathcal P_{\alpha_1\to\omega}(v_1-v_2)+\left(\mathcal P_{\alpha_1\to\omega}-\mathcal P_{\alpha_2\to\omega}\right)(v_2).\]
Since parallel transport is an isometry, $\|\mathcal P_{\alpha_1\to\omega}(v_1-v_2)\|=\|v_1-v_2\|$. Moreover, the map $(\alpha,\omega)\mapsto \mathcal P_{\alpha\to\omega}$ is smooth on $\mathcal S^{d-1}_+\times\mathcal S^{d-1}_+$ and, by compactness, uniformly Lipschitz in $\alpha$. Therefore,
\[\bigl\|\left(\mathcal P_{\alpha_1\to\omega}-\mathcal P_{\alpha_2\to\omega}\right)(v_2)\bigr\|\le Cd_S(\alpha_1,\alpha_2)\|v_2\|\le Cd_S(\alpha_1,\alpha_2).\]
Finally, the logarithmic map $(\alpha,\beta)\mapsto \mathrm{Log}_\alpha(\beta)$ is smooth on $\mathcal S^{d-1}_+\times\mathcal S^{d-1}_+$ and hence uniformly Lipschitz on this compact set, yielding
\[\|v_1-v_2\|\le C\{d_S(\alpha_1,\alpha_2)+d_S(\beta_1,\beta_2)\}.\]
Combining the above bounds yields
\[d_S(\Gamma_{\alpha_1,\beta_1}(\omega),\Gamma_{\alpha_2,\beta_2}(\omega))\le C_2\{d_S(\alpha_1,\alpha_2)+d_S(\beta_1,\beta_2)\}\]
for some constant $C_2>0$, verifying Assumption~4.

\medskip
\noindent\textbf{Verification of Assumption~5.}
Since $\mathcal S^{d-1}_+$ is a compact, geodesically convex subset of the unit sphere equipped with the intrinsic metric, the existence and uniqueness of population and empirical Fr\'echet means, the covering number condition, and the local quadratic growth of the Fr\'echet function follow from standard results for Fr\'echet means on manifolds; see Proposition~3 of \citet{mull:19:6}. In particular, Assumption~5 holds with $\kappa=2$.
\end{proof}

\section{Proofs}\label{supp:proof}
\subsection{Proof of Theorem~1 in the main paper}
\begin{proof}
Observe that $\nu_{0,0}(0)=\nu_{0,0}$ and $\nu_{0,1}(0)=\nu_{0,1}$ since we observe $Y_{i,t}(0)$ directly for $t\in\{0, 1\}$ for the control group. For the treated group, we likewise have $\nu_{1,0}(0)=\nu_{1,0}$ and $\nu_{1,1}(1)=\nu_{1,1}$. By Assumption~2 in the main paper (parallel trends), we have $\Gamma_{\nu_{0,0}, \nu_{0,1}}(\nu_{1,0})=\Gamma_{\nu_{0,0}(0), \nu_{0,1}(0)}(\nu_{1,0}(0))=\nu_{1,1}(0)$. It follows that
\begin{align*}
\ominus \gamma_{\nu_{0,0},\nu_{0,1}} \oplus \gamma_{\nu_{1,0},\nu_{1,1}}
&= \gamma_{\Gamma_{\nu_{0,0},\nu_{0,1}}(\nu_{1,0}),\nu_{1,1}}
= \gamma_{\nu_{1,1}(0),\nu_{1,1}}
= \gamma_{\nu_{1,1}(0), \nu_{1,1}(1)}.
\end{align*}

\end{proof}

\subsection{Proof of Proposition~1 in the main paper}
\begin{proof}
    To show that $\sim$ is an equivalence relation on $\mathcal{G}(\mathcal{M})$, we verify reflexivity, symmetry, and transitivity. Reflexivity, $\gamma_{\alpha,\beta}\sim\gamma_{\alpha,\beta}$, follows directly from the geodesic transport map in Assumption~1 of the main paper. Symmetry and transitivity follow from the definition of $\sim$.
\end{proof}

\subsection{Proof of Proposition~2 in the main paper}
\begin{proof}
To demonstrate that the quotient space of geodesics $\mathcal{G}(\mathcal{M})/\sim$ is a metric space with the metric
\[d_{\mathcal{G}}([\gamma_{\alpha_1, \beta_1}], [\gamma_{\alpha_2, \beta_2}]) = d(\Gamma_{\alpha_1, \beta_1}(\omega_\oplus), \Gamma_{\alpha_2, \beta_2}(\omega_\oplus)),\]
we verify positivity, symmetry, and the triangle inequality.

\noindent\textbf{Positivity:}

\noindent By definition of the metric, $d_{\mathcal{G}}([\gamma_{\alpha_1, \beta_1}], [\gamma_{\alpha_2, \beta_2}]) \geq 0$. If $\gamma_{\alpha_1, \beta_1} \sim \gamma_{\alpha_2, \beta_2}$, then $\Gamma_{\alpha_1, \beta_1}(\omega) = \Gamma_{\alpha_2, \beta_2}(\omega)$ for all $\omega\in\mathcal{M}$, and thus:
\[d_{\mathcal{G}}([\gamma_{\alpha_1, \beta_1}], [\gamma_{\alpha_2, \beta_2}]) = d(\Gamma_{\alpha_1, \beta_1}(\omega_\oplus), \Gamma_{\alpha_2, \beta_2}(\omega_\oplus)) = 0.\]
Conversely, if $d_{\mathcal{G}}([\gamma_{\alpha_1, \beta_1}], [\gamma_{\alpha_2, \beta_2}]) = 0$, then:
\[d(\Gamma_{\alpha_1, \beta_1}(\omega_\oplus), \Gamma_{\alpha_2, \beta_2}(\omega_\oplus)) = 0,\]
which implies $\Gamma_{\alpha_1, \beta_1}(\omega) = \Gamma_{\alpha_2, \beta_2}(\omega)$  for all $\omega\in\mathcal{M}$, and therefore $\gamma_{\alpha_1, \beta_1} \sim \gamma_{\alpha_2, \beta_2}$.

\noindent\textbf{Symmetry:}

\noindent By definition of the metric,
\[d_{\mathcal{G}}([\gamma_{\alpha_1, \beta_1}], [\gamma_{\alpha_2, \beta_2}]) = d(\Gamma_{\alpha_1, \beta_1}(\omega_\oplus), \Gamma_{\alpha_2, \beta_2}(\omega_\oplus)).\]
Since the metric $d$ on $\mathcal{M}$ is symmetric, we have:
\[d(\Gamma_{\alpha_1, \beta_1}(\omega_\oplus), \Gamma_{\alpha_2, \beta_2}(\omega_\oplus)) = d(\Gamma_{\alpha_2, \beta_2}(\omega_\oplus), \Gamma_{\alpha_1, \beta_1}(\omega_\oplus)),\]
and thus:
\[d_{\mathcal{G}}([\gamma_{\alpha_1, \beta_1}], [\gamma_{\alpha_2, \beta_2}]) = d_{\mathcal{G}}([\gamma_{\alpha_2, \beta_2}], [\gamma_{\alpha_1, \beta_1}]).\]

\noindent\textbf{Triangle inequality:}

\noindent Using the definition of the metric,
\[d_{\mathcal{G}}([\gamma_{\alpha_1, \beta_1}], [\gamma_{\alpha_2, \beta_2}]) = d(\Gamma_{\alpha_1, \beta_1}(\omega_\oplus), \Gamma_{\alpha_2, \beta_2}(\omega_\oplus)).\]
By the triangle inequality for $d$ on $\mathcal{M}$, we have:
\[d(\Gamma_{\alpha_1, \beta_1}(\omega_\oplus), \Gamma_{\alpha_2, \beta_2}(\omega_\oplus)) \leq d(\Gamma_{\alpha_1, \beta_1}(\omega_\oplus), \Gamma_{\alpha_3, \beta_3}(\omega_\oplus)) + d(\Gamma_{\alpha_3, \beta_3}(\omega_\oplus), \Gamma_{\alpha_2, \beta_2}(\omega_\oplus)).\]
Substituting back into the metric on the quotient space, this becomes:
\[d_{\mathcal{G}}([\gamma_{\alpha_1, \beta_1}], [\gamma_{\alpha_2, \beta_2}]) \leq d_{\mathcal{G}}([\gamma_{\alpha_1, \beta_1}], [\gamma_{\alpha_3, \beta_3}]) + d_{\mathcal{G}}([\gamma_{\alpha_3, \beta_3}], [\gamma_{\alpha_2, \beta_2}]).\]

The quotient space $\mathcal{G}(\mathcal{M})/\sim$ with the metric
\[d_{\mathcal{G}}([\gamma_{\alpha_1, \beta_1}], [\gamma_{\alpha_2, \beta_2}]) = d(\Gamma_{\alpha_1, \beta_1}(\omega_\oplus), \Gamma_{\alpha_2, \beta_2}(\omega_\oplus))\]
satisfies positivity, symmetry, and triangle inequality, and is therefore a metric space.
\end{proof}

\subsection{Proof of Theorem~2 in the main paper}
\begin{proof}
By the definition of $d_{\mathcal{G}}$ and Assumption~4 in the main paper, we have
\begin{align*}
d_{\mathcal{G}}(\hat{\tau},\tau)&=d_{\mathcal{G}}(\gamma_{\hat{\nu}_{1,1}', \hat{\nu}_{1,1}}, \gamma_{\nu_{1,1}', \nu_{1,1}})\\&=d(\Gamma_{\hat{\nu}_{1,1}', \hat{\nu}_{1,1}}(\nu_{1,1}'), \Gamma_{\nu_{1,1}', \nu_{1,1}}(\nu_{1,1}'))\\&\leq C_2\{d(\hat{\nu}_{1,1}', \nu_{1,1}')+d(\hat{\nu}_{1,1}, \nu_{1,1})\}.
\end{align*}
For the first term $d(\hat{\nu}_{1,1}', \nu_{1,1}')$, applying the triangle inequality gives
\begin{align*}
    d(\nu_{1,1}', \hat{\nu}_{1,1}')&=d(\Gamma_{\nu_{0,0},\nu_{0,1}}(\nu_{1,0}), \Gamma_{\hat{\nu}_{0,0},\hat{\nu}_{0,1}}(\hat{\nu}_{1,0}))\\&\leq d(\Gamma_{\nu_{0,0},\nu_{0,1}}(\nu_{1,0}),\Gamma_{\nu_{0,0},\nu_{0,1}}(\hat{\nu}_{1,0}))+d(\Gamma_{\nu_{0,0},\nu_{0,1}}(\hat{\nu}_{1,0}), \Gamma_{\hat{\nu}_{0,0},\hat{\nu}_{0,1}}(\hat{\nu}_{1,0})).
\end{align*}
Using Assumptions~3 and~4 in the main paper, we further obtain
\[d(\nu_{1,1}', \hat{\nu}_{1,1}')\leq C_1d(\hat{\nu}_{1,0}, \nu_{1,0})+C_2\{d(\hat{\nu}_{0,0}, \nu_{0,0})+d(\nu_{0,1},\hat{\nu}_{0,1})\}.\]
Combining these results, we have
\[d_{\mathcal{G}}(\hat{\tau},\tau)\leq C_1C_2d(\hat{\nu}_{1,0}, \nu_{1,0})+C_2^2d(\hat{\nu}_{0,0}, \nu_{0,0})+C_2^2d(\nu_{0,1},\hat{\nu}_{0,1})+C_2d(\hat{\nu}_{1,1}, \nu_{1,1}).\]
Under Assumption~5 in the main paper and by Theorem~2 of \citet{mull:19:6}, one has
\[d(\hat{\nu}_{d, t}, \nu_{d, t})=O_p(n^{-1/(2(\kappa-1))}),\quad d, t\in\{0, 1\}.\]
Substituting these convergence rates into the above inequality, we conclude
\[d_{\mathcal{G}}(\hat{\tau},\tau)=O_p(n^{-1/(2(\kappa-1))}).\]
\end{proof}

\subsection{Proof of Theorem~\ref{thm:GDD-stag-id}}
\begin{proof}
We first show  (i). Observe that for $s= g-\delta-1$,
\begin{align*}
\beta_{s+1}
&= \Gamma_{\Eo\{Y_{i,s}\mid C_i=1\},\Eo\{Y_{i,s+1}\mid C_i=1\}}(\beta_{s})\\
&= \Gamma_{\Eo\{Y_{i,s}(0)\mid C_i=1\},\Eo\{Y_{i,s+1}(0)\mid C_i=1\}}(\Eo\{Y_{i,s}(0)\mid G_{i,g}=1\})\\
&= \Eo\{Y_{i,s+1}(0)\mid G_{i,g}=1\},
\end{align*}
where the second equality follows from Assumptions~\ref{ass:irrev-treat} and~\ref{ass:lim-t-anti}, and the third equality follows from Assumption~\ref{ass:pt-nev}. Likewise, we have $\beta_s = \Eo\{Y_{i,s}(0)\mid G_{i,g}=1\}$, $s=g-\delta,\ldots,t$. Then we have
\begin{align*}
\gamma_{\beta_t, \Eo\{Y_{i,t}\mid G_{i,g}=1\}}
&= \gamma_{\Eo\{Y_{i,t}(0)\mid G_{i,g}=1\}, \Eo\{Y_{i,t}\mid G_{i,g}=1\}}\\
&= \gamma_{\Eo\{Y_{i,t}(0)\mid G_{i,g}=1\}, \Eo\{Y_{i,t}(g)\mid G_{i,g}=1\}}\\
&= \tau(g,t),
\end{align*}
where the second equality follows from the definition of potential outcomes.

Next, we show (ii). Observe that for $s= g-\delta-1$,
\begin{align*}
\beta_{s+1}
&= \Gamma_{\nu_s,\nu_{s+1}}(\beta_s)\\
&= \Gamma_{\nu_s,\nu_{s+1}}(\Eo\{Y_{i,s}(0)\mid G_{i,g}=1\})\\
&= \Eo\{Y_{i,s+1}(0)\mid G_{i,g}=1\},
\end{align*}
where the second equality follows from Assumptions~\ref{ass:irrev-treat} and~\ref{ass:lim-t-anti}, and the third equality follows from Assumption~\ref{ass:pt-ny}. Likewise, we have $\beta_s = \Eo\{Y_{i,s}(0)\mid G_{i,g}=1\}$, $s=g-\delta,\ldots,t$. Then we have
\begin{align*}
\gamma_{\beta_t, \Eo\{Y_{i,t}\mid G_{i,g}=1\}}
&= \gamma_{\Eo\{Y_{i,t}(0)\mid G_{i,g}=1\}, \Eo\{Y_{i,t}\mid G_{i,g}=1\}}\\
&= \gamma_{\Eo\{Y_{i,t}(0)\mid G_{i,g}=1\}, \Eo\{Y_{i,t}(g)\mid G_{i,g}=1\}}\\
&= \tau(g,t),
\end{align*}
where the second equality follows from the definition of potential outcomes.
\end{proof}

\subsection{Proof of Corollary~\ref{cor:GDD-stag-id}}
\begin{proof}
We only show (i), as (ii) follows in the same way. Observe that
\begin{align*}
\Eo\{Y_{i,t}(0)\mid G_{i,g}=1\}
&=\beta_t \\
&= \Gamma_{\alpha_{t-1},\alpha_t}(\beta_{t-1})\\
&\ \ \vdots \\
&= (\Gamma_{\alpha_{t-1},\alpha_t} \circ \cdots \circ \Gamma_{\alpha_{g-\delta-1},\alpha_{g-\delta}})(\beta_{g-\delta-1})\\
&= \Gamma_{\alpha_{g-\delta-1},\alpha_t}(\beta_{g-\delta-1}),
\end{align*}
where $\alpha_s = \Eo\{Y_{i,s}(0)\mid C_i=1\}$, $s=g-\delta-1,\ldots,t$. Then we obtain
\begin{align*}
&\ominus \gamma_{\Eo\{Y_{i,g-\delta-1}\mid C_i=1\}, \Eo\{Y_{i,t}\mid C_i=1\}} \oplus \gamma_{\Eo\{Y_{i,g-\delta-1}\mid G_{i,g}=1\}, \Eo\{Y_{i,t}\mid G_{i,g}=1\}}\\
&= \gamma_{\Gamma_{\alpha_{g-\delta-1},\alpha_t}(\beta_{g-\delta-1}), \Eo\{Y_{i,t}(g)\mid G_{i,g}=1\}}\\
&= \gamma_{\beta_t,\Eo\{Y_{i,t}(g)\mid G_{i,g}=1\}}\\
&= \tau(g,t),
\end{align*}
where the first equality follows from the definition of geodesic difference and consistency of observed and potential outcomes.
\end{proof}

\end{document}